\documentclass[aps,prd,floatfix,nofootinbib,showpacs,superscriptaddress]{revtex4}

\usepackage{graphicx,amsmath,amssymb,bm}
\usepackage[english]{babel}
\usepackage{color}
\usepackage[utf8]{inputenc}
\definecolor{purple}{rgb}{0.5,0,0.5}
\definecolor{blue}{rgb}{0.0,0,0.9}
\usepackage[colorlinks=true, pdfstartview=FitV, citecolor= purple, linkcolor = blue,urlcolor=blue]{hyperref}
\usepackage{amssymb}
\usepackage{graphics}
\usepackage{epsfig}
\usepackage{rotating}
\usepackage{longtable} 
\usepackage{url}
\usepackage{pdflscape}
\usepackage{cancel}
\pdfoutput=1

\usepackage{array}
\usepackage{multirow}
\newcolumntype{L}[1]{>{\raggedright\arraybackslash}p{#1}}
\newcolumntype{C}[1]{>{\centering\arraybackslash}p{#1}}
\newcolumntype{R}[1]{>{\raggedleft\arraybackslash}p{#1}}

\newcommand{\gmor}{^\textrm{\tiny GMOR}}

\begin{document}

\title{The Podolsky propagator in gap and bound-state equations}

\author{Bruno El-Bennich}
 \email{bruno.bennich@cruzeirodosul.edu.br}
 \affiliation{Laborat\'orio de F\'isica Te\'orica e Computacional, Universidad Cidade de S\~ao Paulo, Rua Galv\~ao Bueno 868, 01506-000, S\~ao Paulo, SP, Brazil}
\author{German Ramos-Zambrano}
\email{gramoszge@gmail.com}
 \affiliation{Departamento de F\'\i sica, Universidad de Nari\~no, A.A. 1175,  San Juan de Pasto, Colombia}
\author{Eduardo Rojas}
\email{eduro4000@gmail.com}
\affiliation{Departamento de F\'\i sica, Universidad de Nari\~no, A.A. 1175, San Juan de Pasto, Colombia}
%

\begin{abstract}
Based on the Generalized Quantum Electrodynamics expression for the Podolsky propagator, which preserves gauge invariance for massive photons, we propose a model for the massive gluon 
propagator that reproduces well-known features of established strong-interaction models in the framework of the Dyson-Schwinger equation. By adjusting the Podolsky mass and the coupling 
strength we thus construct a model with simple analytical properties known from perturbative theory, yet well suited to describe a confining interaction. We obtain solutions of the Dyson-Schwinger 
equation for the quark at space-like momenta on the real axis as well as on the complex plane and solving the bound-state problem with the Bethe-Salpeter equation yields masses and 
weak decay constants of the $\pi, K$ and $\eta_c$ in excellent agreement with experimental values, while the $D$ and $D_s$ are  reasonably well described. The analytical simplicity of 
this effective interaction has the potential to be useful for  phenomenological  applications and may facilitate calculations in Minkowski space. 
\end{abstract}

\maketitle


\section{Introduction}

Mandelstam's seminal work~\cite{Mandelstam:1979xd} established that the rainbow-ladder truncation of the gap and bound-state equations is an adequate approximation to describe dynamical 
chiral symmetry breaking (DCSB)~\cite{Cornwall:1974vz,Cornwall:1981zr,Cornwall:1989gv,Burden:1993gy,Brown:1988bn,Fischer:2004nq,Bashir:2005wt,Chang:2009ae,Bashir:2013zha,Bashir:2012fs,
Cloet:2013jya}  in Quantum Chromodynamics (QCD). Later on, within this same truncation scheme, Munczek and Nemirovsky succeeded in reproducing the masses of pseudoscalar and vector 
meson ground states~\cite{Munczek:1983dx}. More sophisticated models succeeded in the following decades that satisfy theoretical and phenomenological constraints~\cite{Pelaez:2017bhh,
Munczek:1988er,Praschifka:1989fd,Williams:1989tv,vonSmekal:1991fp,Jain:1993qh,Frank:1995uk,Maris:1997tm,Maris:1997hd,Maris:1999nt,Alkofer:2002bp,Qin:2011dd}. 
Their popularity owes to a wide range of successful applications to mesons, baryons, hyperons, their excited states and parity partners~\cite{Maris:1997tm,Maris:1997hd,Maris:1999nt,Alkofer:2002bp,
Qin:2011dd,Chang:2011ei,Chang:2013pq,Chang:2013nia,Rojas:2014aka,Raya:2015gva,El-Bennich:2016qmb,El-Bennich:2016bno,Mojica:2017tvh,Shi:2018zqd,Qin:2020jig,Serna:2020txe,Cloet:2008re,
Eichmann:2009qa,Aznauryan:2012ba,Segovia:2015hra,Eichmann:2016hgl,Eichmann:2016yit,Chen:2017pse,Sanchis-Alepuz:2017jjd,Chen:2018nsg,Bednar:2018htv,Qin:2019hgk,Chen:2019fzn}.

In this work we propose a model that describes effectively a massive gluon interaction in the infrared region, where we are inspired by the functional  structure of Generalized Quantum Electrodynamics~(GQED) 
proposed long ago by Podolsky~\cite{Podolsky:1942zz,Podolsky:1944zz}. Historically, this generalization aimed at remedying pathologies inherent to the Maxwell theory and consisted in introducing 
higher-order derivatives in the Lagrangian of electrodynamics, maintaining at the same time linearity of the equations of motion in the fields. In other words, the goal was to eliminate the infinities that 
arise in higher-order corrections of point charges and the associated coupling. 

However, since this extension of the Lagrangian preserves gauge invariance in a consistent treatment~\cite{Galvao:1986yq,Bufalo:2012tt}, GQED has come to be viewed more as a prototype 
of a theory that contains massless \emph{as well as\/} massive photons that do not break gauge invariance. This is because Podolsky's extension of electrodynamics is the only possible linear,
Lorentz and $U(1)$ invariant generalization of the Maxwell theory~\cite{Cuzinatto:2005zr} and a consistent quantization of GQED was shown to require a generalized Landau gauge 
condition~\cite{Galvao:1986yq}, while the proper covariant quantization of GQED in this generalized gauge was obtained with functional methods in Ref.~\cite{Bufalo:2010sb}. More recently, 
the Podolsky approach to QED was also reinterpreted as a natural way of providing a Pauli-Villars regularization in ordinary QED~\cite{Ji:2019phv}. GQED introduces therefore in a consistent
manner a mass parameter $m_P$ in the vector-boson propagator while preserving gauge invariance and acting as an effective ultraviolet cutoff in Landau gauge. 

These features are clearly attractive for modeling the nonperturbative gluon interaction in an Abelianized truncation of QCD, given the compelling body of work that evidence an infrared-finite gluon 
propagator~\cite{Fischer:2008uz,Alkofer:2008jy,Dudal:2008sp,Aguilar:2004sw,Aguilar:2008xm,Aguilar:2012rz,Cucchieri:2007md,Cucchieri:2007rg,Oliveira:2008uf,Pennington:2011xs,Oliveira:2012eh,
Bogolubsky:2009dc,Ayala:2012pb,Strauss:2012dg,Huber:2015ria,Cyrol:2016tym,Boucaud:2018xup,Mintz:2018hhx,Dudal:2018cli,Aguilar:2019uob,Gunkel:2019xnh,Gunkel:2020wcl,Huber:2020keu}
and which can be related to an effective gluon mass. It turns out that in Landau gauge and in the  leading truncation of the quark's Dyson-Schwinger equation (DSE) we may interpret the Podolsky 
propagator as a nonperturbative model for the gluon propagator, at least in the low-momentum region, where its  massive  ``dressing function" effectively drives the strength of the DCSB. 

In Section~\ref{DSEsec} we introduce the DSE that describes the quark-gap equation with a Podolsky propagator in Landau gauge and obtain its solutions for different flavors on the space-like 
real axis. The functional behavior of the quark's mass and wave-renormalization function is reminiscent of that found with the Maris-Tandy~\cite{Maris:1999nt} or 
Qin-Chang~\cite{Qin:2011dd}  models and the obvious question arises whether this interaction is useful for hadron phenomenology. We solve the Bethe-Salpeter equation (BSE) as 
usual in Euclidean space to find antiquark-quark bound states, which implies that the arguments of the quark propagators are complex-valued momenta. To obtain the quark propagators 
on the complex plane, we apply Cauchy's integral  theorem that requires DSE solutions on a contour defined by a parabola describing the complex-momentum distribution 
\cite{Fischer:2005en,Krassnigg:2009gd}. We note that the convergence of the DSE on such a contour is not generally guaranteed for a given interaction regardless of its convergence on 
the real axis. Nonetheless, using the Podolsky propagator with an appropriate parameterizations the DSE converges rapidly on this contour and the BSE solutions reproduce the mass spectrum 
and weak decay constants of pseudoscalar mesons as we discuss in Section~\ref{sec:results}.  We finish with some concluding remarks about possible extensions and applications of the 
Podolsky propagator in Section~\ref{conclude}.


\section{Dyson-Schwinger Equation \label{DSEsec} }

The gap equation for a quark of flavor $f$ is expressed by a DSE for the inverse propagator in Minkowski space as,
\begin{equation}
  S_f^{-1}( p)  =   Z_2\, \gamma \cdot p -   Z_4\, m_f(\mu)   -  Z_1\, g^{2} \! \int^\Lambda\! \!\frac{ d^{4}k}{( 2\pi )^{4}} \; G^{\mu \nu}(q)
                                             \frac{\lambda^a}{2} \gamma _{\mu }\, S_f( k) \frac{\lambda^a}{2} \Gamma_{\nu }\left( k,p\right) \ ,
\label{1}
\end{equation}
where $Z_2(\mu,\Lambda )$, $Z_4(\mu,\Lambda )$  and  $Z_1(\mu,\Lambda )$ are the wave-function, mass and vertex renormalization constants, respectively. 
Moreover, $\Gamma^a_\mu (k,p) = \frac{1}{2}\,\lambda^a \Gamma_\mu (k,p) $ is the quark-gluon vertex and $\lambda^a$ are the SU(3) color matrices 
in the fundamental  representation, while  $\Lambda$ is a Poincar\'e-invariant regularization scale, chosen such that $\Lambda \gg \mu$. 
In GQED the vector-boson propagator in a covariant gauge with gauge-fixing parameter $\xi$ and momentum, $q = k-p$, is given by,
\begin{eqnarray}
    G_{\mu \nu } &  =    &  -i \Delta(q^2) P_{\mu \nu }( q)  \  , \nonumber  \\
   P_{\mu \nu }( q) 
        &  =   &   \Delta_{\mu \nu }( q) -\left[ ~g_{\mu \nu }\smallskip +\left( 1-\xi \right) \frac{q_{\mu }q_{\nu }}{q^{2}-m_{P}^{2}}\right] 
         \frac{1}{ \smallskip \smallskip q^{2}-m_{P}^{2}}+\left( \smallskip 1-2\xi \right)  \frac{q_{\mu }q_{\nu }}{q^{2}\left( q^{2}-m_{P}^{2}\right) }~+
         \frac{q_{\mu }q_{\nu }}{\left( q^{2}-m_{P}^{2}\right) ^{2}} \ , \label{podolskyprop} 
\end{eqnarray}
with the standard gauge-boson propagator, 
 \begin{equation}
  \Delta_{\mu \nu } (q)  =\left[ g_{\mu \nu } -   \left( 1-\xi \right) \frac{q_{\mu }q_{\nu }}{q^{2}}\right] \frac{1}{q^{2}} \ .
 \end{equation}
 
In general, the Dirac structure of the fermion propagator is fully defined by two covariants and associated scalar functions, the wave-function 
renormalization $\mathcal{F}_f( p)$ and the mass function $ \mathcal{M}_f( p)$, so that,
\begin{equation}
   S_f ( p ) = \frac{\mathcal{F}_f  (p ) } {\gamma \cdot p -  \mathcal{M}_f ( p ) } \ .
\end{equation}%
In order to determine the renormalization constants and to make quantitative matching with pQCD, one imposes the renormalization conditions,
\begin{equation}
    \left. \mathcal{F}_f(p^2)\right |_{p^2 = \mu^2} = \ 1 \ ,   
\qquad
   \left.  S^{-1}_f (p) \right |_{p^2=\mu^2}  =  \  \gamma\cdot p \ - m_f(\mu )  \ ,
\end{equation}
where  $\mu^2\gg \Lambda_\mathrm{QCD}^2$ and $m_f(\mu )$ is the renormalized running quark mass; in particular, $m_f(\mu )$ 
is nothing else but the dressed-quark mass function evaluated at one particular deep space-like point, $p^2=\mu^2$, namely $m_f(\mu)  =   \mathcal{M}_f  (\mu )$.

\begin{figure}[t!] 
\centering 
 \includegraphics[scale=0.31]{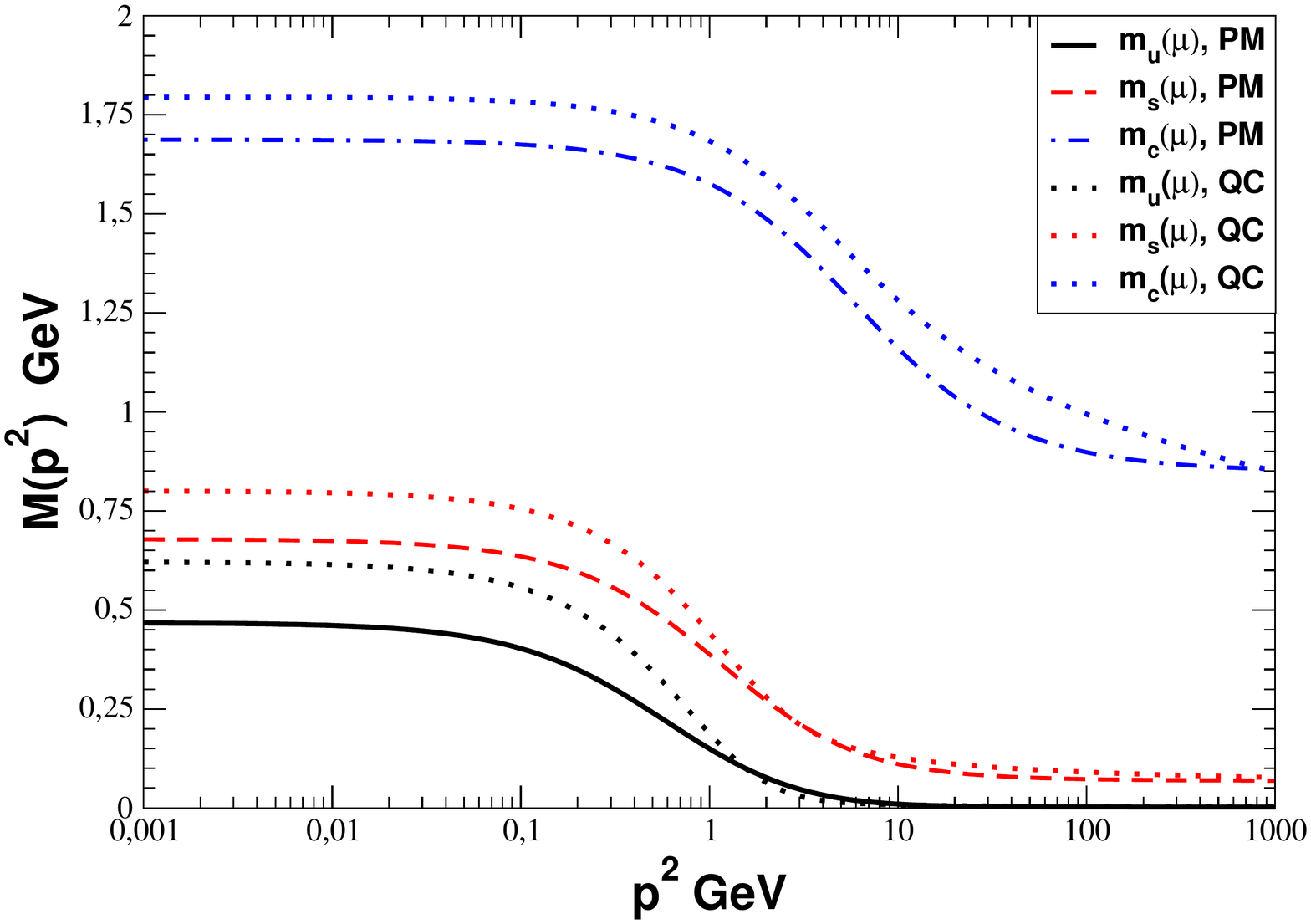}  \hfill \includegraphics[scale=0.31]{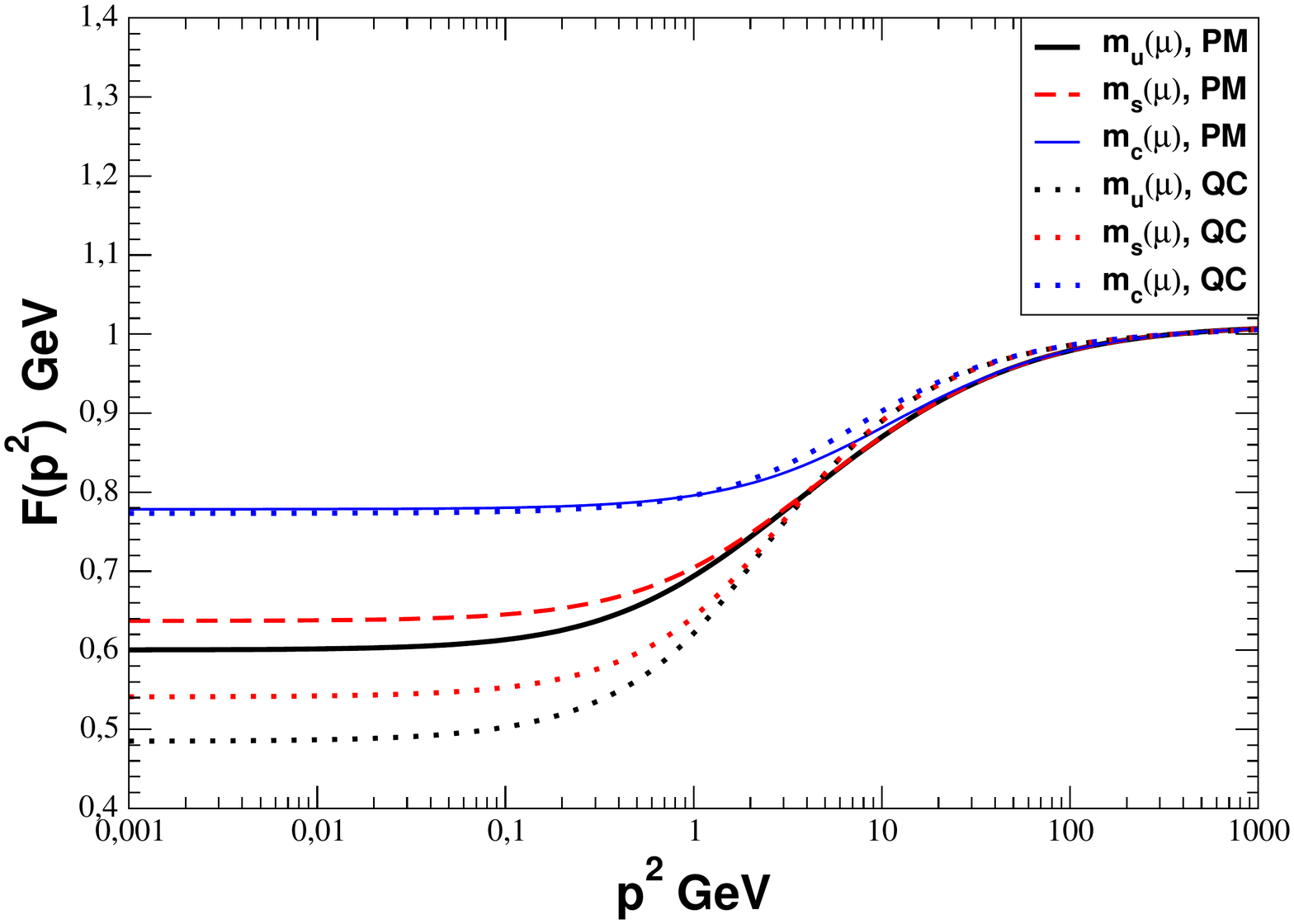}  
 \vspace*{-2mm}
\caption{ Mass function, $ \mathcal{M}_f(p^2)$, and wave function renormalization, $\mathcal{F}_f(p^2)$, for the light, strange and charm quarks obtained with  
            Eqs.~\eqref{eq:sde1} and \eqref{eq:sde2} using the Podolsky model (PM)~\eqref{eq:g2} compared to DSE solutions calculated with the model in Ref.~\cite{Qin:2011dd} (QC). 
            All functions are renormalized at $\mu =19$~GeV with $m_{u,d}(19\,\mathrm{GeV}) =2.64$~MeV,  $m_s (19\,\mathrm{GeV}) = 70$~MeV and  $m_c (19\,\mathrm{GeV}) = 865$~MeV. }
  \label{masswavefunc}
\end{figure}

The mass function $\mathcal{M}_f(p^2)$, and the the renormalization wave function $\mathcal{F}_f$  can be projected out from the DSE~\eqref{1}  and rewritten in Euclidean 
space one obtains\footnote{ See Appendix~\ref{sec:appendixA} for details of the calculation in arbitrary covariant gauge.} 
in Landau gauge and for a bare vertex, $\Gamma_\nu (k,p) = \gamma_\nu$, the two coupled, nonlinear integral equations,
\begin{eqnarray}
    \frac{ \mathcal{M}_f ( p_{E} ) }{\mathcal{F}_f ( p_{E} ) } &=  & Z_4 m(\mu) + 3C_F\int \frac{d^{4}k_{E}}{( 2\pi ) ^{4}}
    \frac{ \mathcal{M}_f ( k_{E} ) \mathcal{F}_f ( k_{E} ) }{k_{E}^{2}+ \mathcal{M}_f^{2} ( k_{E} ) } \frac{\mathcal{G}(q^2_E)}{q_{E}^{2}} \ ,
\label{eq:sde1}
  \\
    \frac{1}{\mathcal{F}_f ( p_{E} ) }  & = &  Z_2+\frac{C_F}{p_{E}^{2}}\int \frac{d^{4}k_{E}}{( 2\pi ) ^{4}} \frac{\mathcal{F}_f(k_{E} ) }
      {k_{E}^{2}+ \mathcal{M}_f^{2} ( k_{E} ) }  \left  [ 3 \,  p_{E}\cdot k_{E}  + \frac{ 2 [ \left( p_{E}\cdot k_{E}\right)^{2}
      - p_{E}^{2}k_{E}^{2} ] }{q_{E}^{2}} \right  ]  \frac{\mathcal{G}(q^2_E)}{q_{E}^{2}} \   .
\label{eq:sde2}
\end{eqnarray}
We define an interaction model by, 
\begin{equation}
     \mathcal{G} (q^2_E) \: = \  4\pi \alpha_{\mathrm{eff}}\,  \frac{ m_{P}^{2} } {q_{E}^{2}+m_{P}^{2} } \ ,
\label{eq:g2}
\end{equation}
with the gluon momentum, $q_E=k_E-p_E$, and the numerical values $\alpha_{\text{eff}} = Z_1g^2/4\pi = 7.69$  and $m_P^2= 0.6$~GeV$^2$. These values have been chosen to reproduce the pion's 
mass and weak decay constant,  as will be discussed in Section~\ref{sec:results}. Note that once fixed, these same parameters are employed for other mesons.  In our approach we set $\Delta(q^2)=1$ 
in Eq.~\eqref{podolskyprop}, that is the perturbative value, such that the gluon-dressing function is defined implicitly by Eq.~\eqref{eq:g2} and thus parameterized by $m_{P}$ and $g^2$.  
As can be appreciated from Fig.~\ref{masswavefunc}, the mass functions, $\mathcal{M}_f$, and wave-renormalization functions, $\mathcal{F}_f$, obtained with this model compare well with the 
functional form that results from the rainbow-ladder model  introduced in Ref.~\cite{Qin:2011dd}, though $\mathcal{F}_f$ is more suppressed with the Podolsky model. This is in particular the case
for the lighter quark flavors.  Likewise, the mass functions are also suppressed in the momentum region $p^2 \lesssim 1$~GeV$^2$.


We note that the Podolsky mass, $m_P\approx 0.77$~GeV, which serves as an infrared mass scale in our model, agrees with other gluon mass scales~\cite{Aguilar:2009nf,Tissier:2010ts,Pelaez:2017bhh}. 
In particular, in Landau gauge, the integral expressions for $\mathcal{M}_f ( p_E )$ and $\mathcal{F}_f ( p_E )$ in Eqs.~\eqref{eq:sde1} and \eqref{eq:sde2} can also be obtained by taking 
the projections and traces of  the DSE~\eqref{1} using a modified perturbative gluon propagator with a mass term: 
\begin{equation}
  D_{\mu \nu } (q) := \frac{m_P^2}{q_E^2+m_P^2}  \left ( g_{\mu \nu } -  \frac{q_{E\mu} q_{E\nu}}{q^2_E} \right ) \frac{1}{q^2_E} \  = \ \frac{m_P^2}{q_E^2+m_P^2} \, \Delta_{\mu\nu}^\mathrm{\xi=0} (q_E^2)\ .
  \label{masspropagator}
\end{equation}
Indeed, for $\xi =0$ and with $\Delta (q^2) =1$ the Podolsky propagator~\eqref{podolskyprop} reduces to Eq.~\eqref{masspropagator} and is equivalent to multiplying by a factor $m_P^2/q_E^2$ 
the massive gluon propagator introduced in Ref.~\cite{Tissier:2010ts}. The latter is motivated by soft breaking of  BRST symmetry and described by a gluon-mass term in the Lagrangian corresponding 
to a particular case of the Curci-Ferrari model~\cite{Curci:1976bt}. Alternatively, here the vector-boson mass $m_P$ originates in higher-order derivatives which preserve gauge symmetry in a generalized 
Lagrangian of the Abelian field theory. The coupling $\alpha_{\text{eff}}$, on the other hand, cannot be directly compared  with the running strong coupling at this scale. This is because we merely 
employ the rainbow truncation of the DSE in which $\alpha_{\text{eff}}$ partially accounts for the lack of DCSB from a fully dressed quark-gluon vertex~\cite{Alkofer:2008tt,Kizilersu:2009kg,Bashir:2011dp,
Rojas:2013tza,Rojas:2014tya,Serna:2018dwk,Aguilar:2014lha,Williams:2014iea,Pelaez:2015tba,Pennington:2016vxv,Williams:2015cvx,Sternbeck:2017ntv,Aguilar:2018epe,Albino:2018ncl,Oliveira:2018ukh,
Oliveira:2020yac,Gao:2020qsj}. 

In this context, we remind that one can infer qualitative and analytic properties of the interaction kernel from the mass and  wave-renormalization functions via an inversion process of the 
DSE~\cite{Rojas:2013tza,Rojas:2014tya} which allows for the comparison of different models. We also stress that it is possible to modify the interaction~\eqref{eq:g2} to include the perturbative 
$\sim 1/q^2$ running at larger momenta. However, our aim here merely consists in verifying the simple expression's~\eqref{eq:g2} capacity to yield the adequate DCSB observed in hadron phenomenology. 

\begin{figure}[t!] 
\centering 
 \includegraphics[scale=0.22]{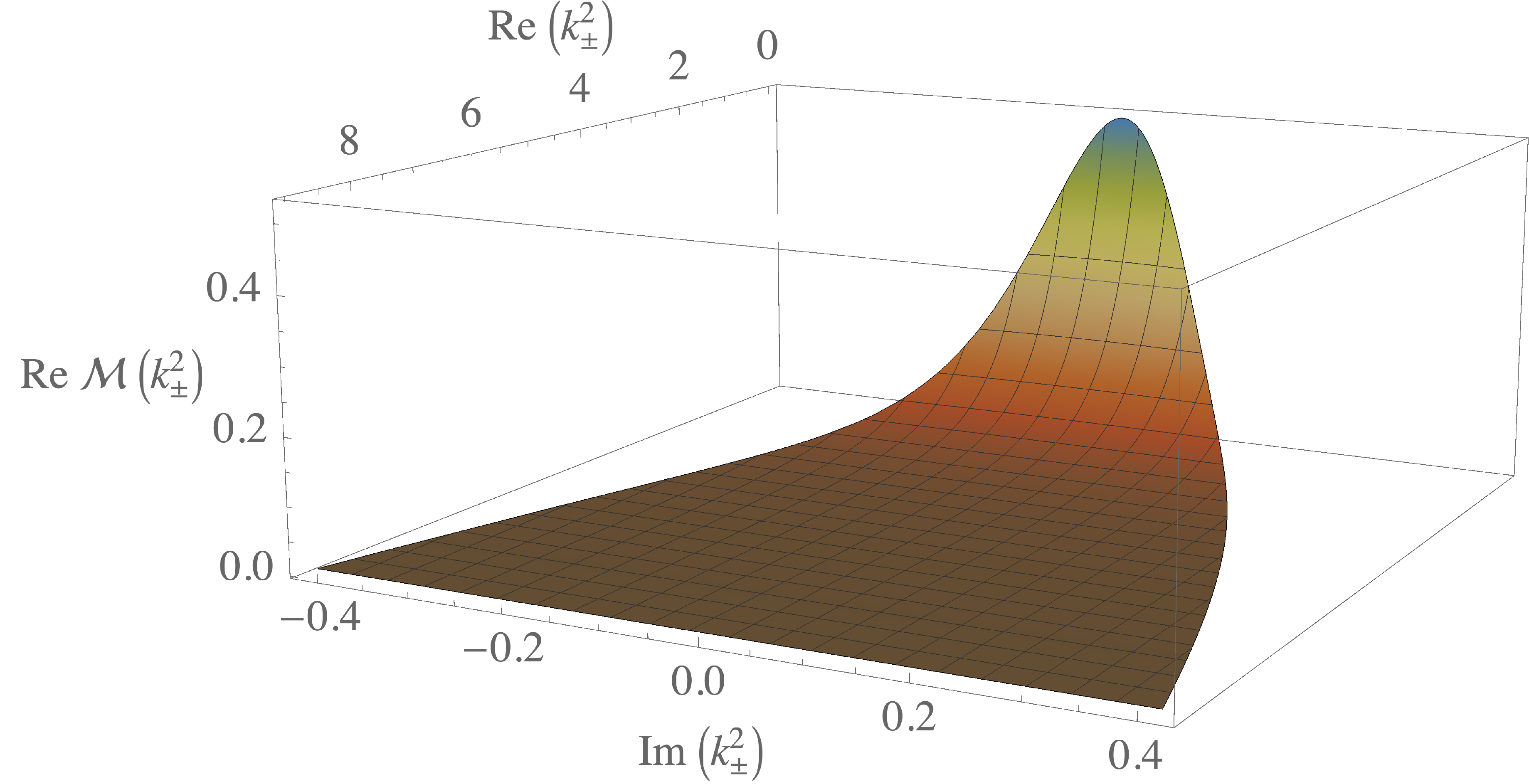}\hfill \includegraphics[scale=0.22]{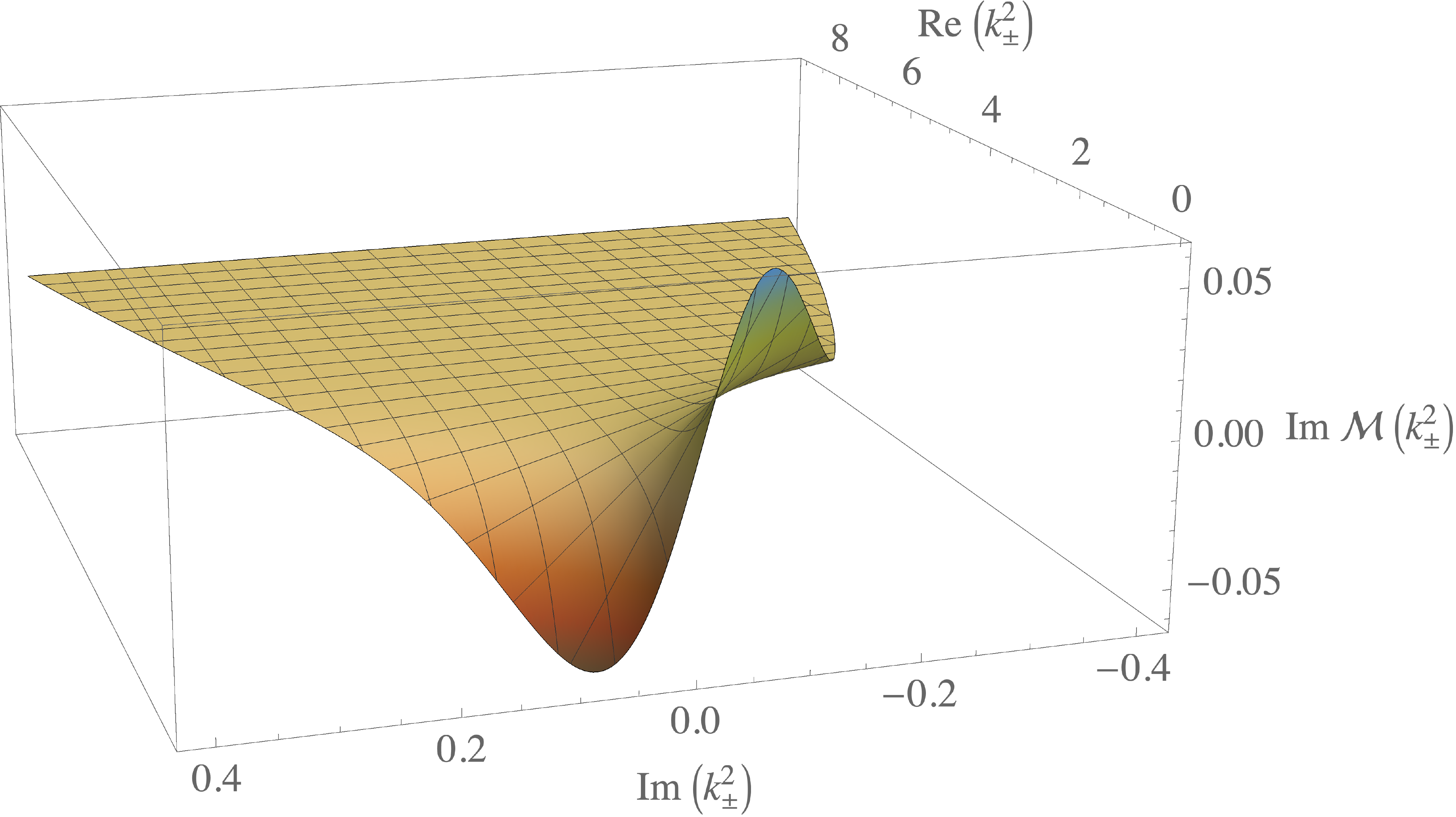}  
\caption{Real part (left panel) and imaginary part (right panel) of the mass function $\mathcal{M}_u(p^2)$ for a current-quark mass $m(19\,\mathrm{GeV}) =2.64$~MeV and an external pion mass 
              $P^2 = -M_\pi^2$  using the Podolsky model~\eqref{eq:g2}. 
              The parabolic area on the complex $k^2_\pm$ plane (in GeV$^2$) is defined by the quark momentum in the BSE for a pion as in Eq.~\eqref{BSEmomentum}.}
  \label{figMReIm}
\end{figure}

\begin{figure}[t!] 
\centering 
 \includegraphics[scale=0.22]{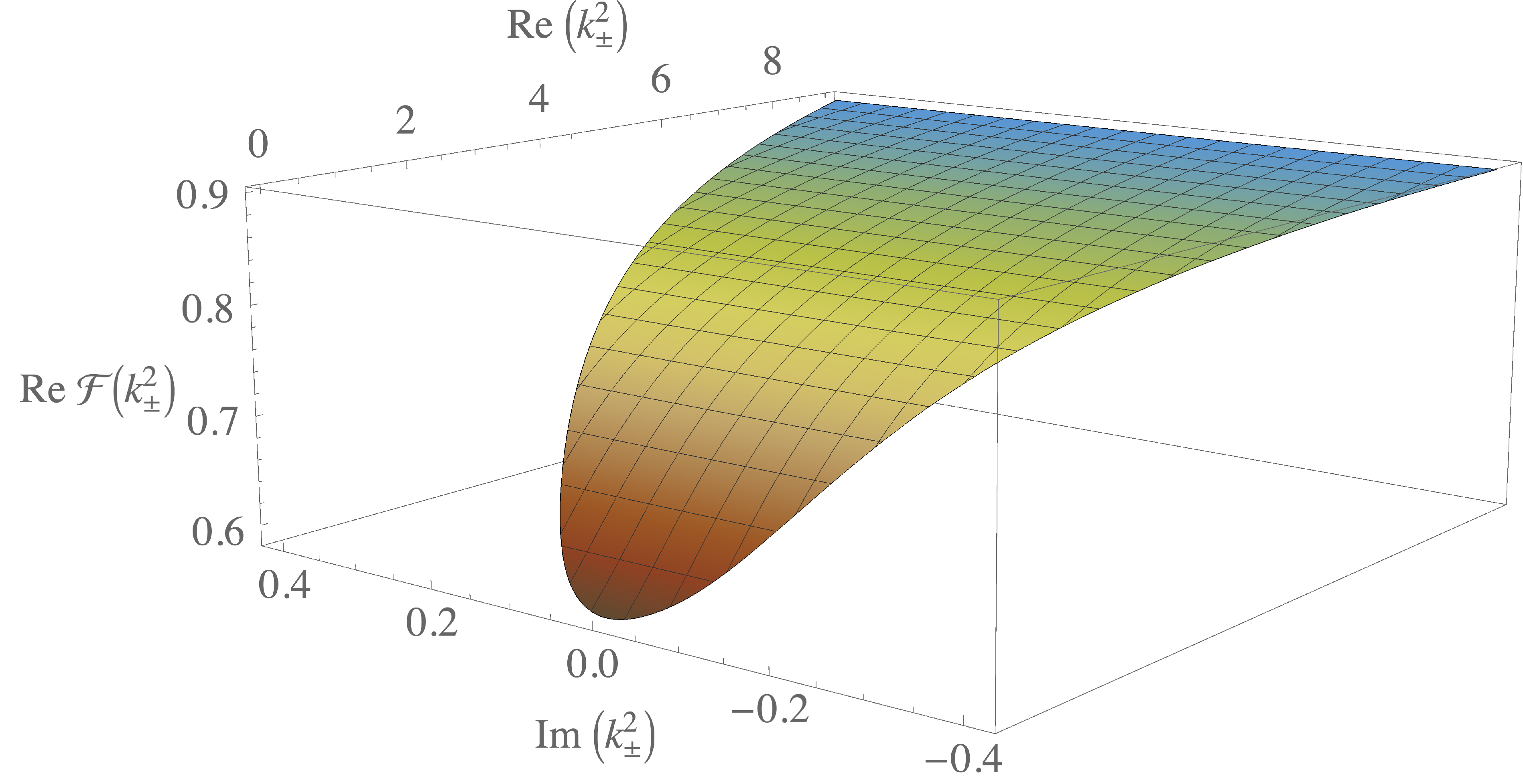}\hfill \includegraphics[scale=0.22]{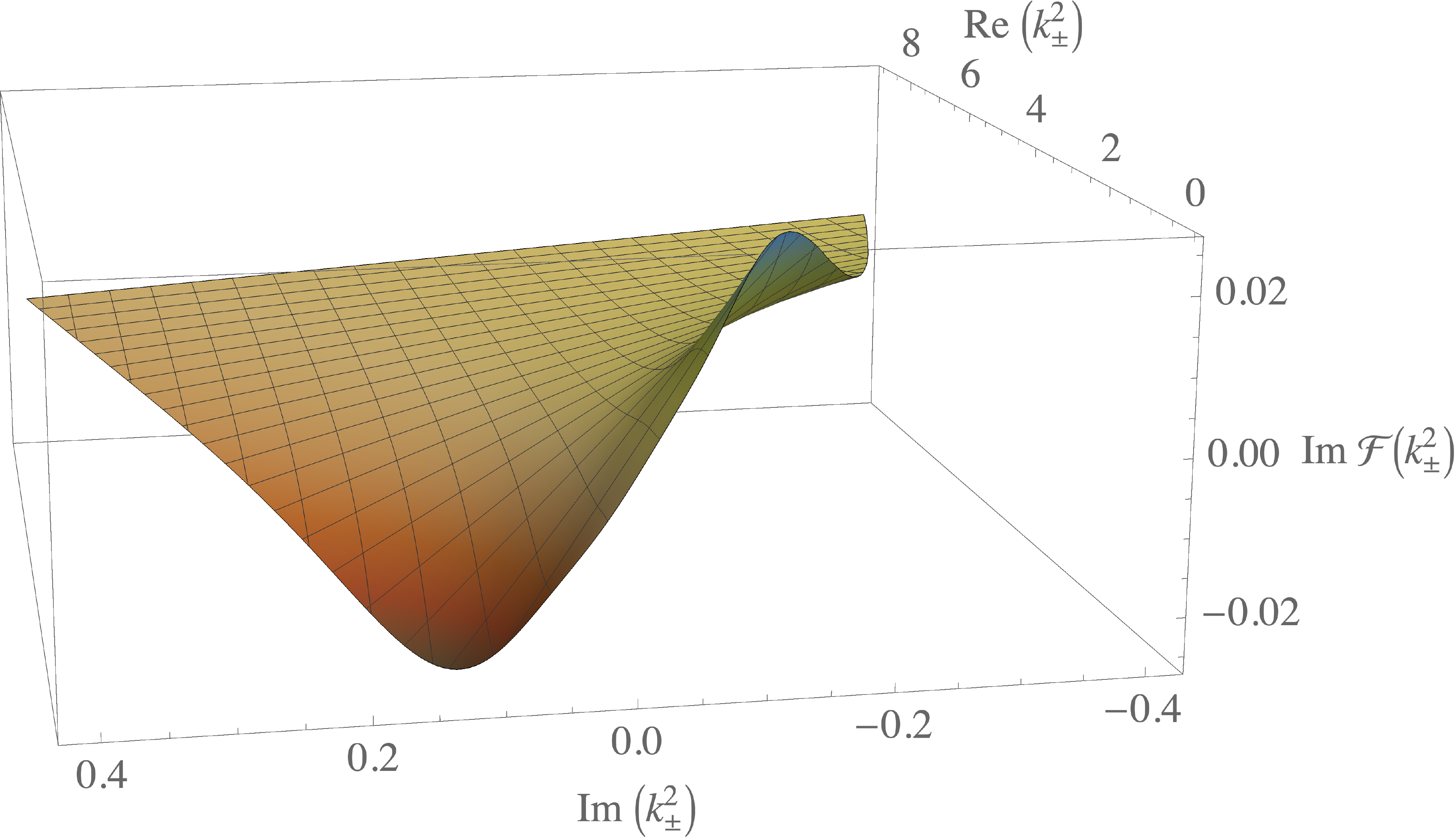}  
\caption{Real part (left panel) and imaginary part (right panel) of the wave-renormalization function $\mathcal{F}_u(p^2)$ for a current-quark mass $m(19\,\mathrm{GeV}) =2.64$~MeV and an external pion mass 
              $P^2 = -M_\pi^2$  using the Podolsky model~\eqref{eq:g2}. }
  \label{figFReIm}
\end{figure}

We close this section with the graphs of $\mathcal{M}_u(k^2)$ and $\mathcal{F}_u(k^2)$ functions on the complex plane, i.e. their solutions calculated on a parabola of complex momenta $p^2$ defined 
by the arguments of the quark propagators in the BSE~\eqref{BSE},
\begin{equation}
   k^2_\pm  =  (k \pm\eta_\pm P)^2 =  k^2  - \eta_\pm^2 M^2_P  \pm 2 i \eta_\pm \, M_P  | k | z_k \  , 
 \label{BSEmomentum}
\end{equation}
where $P^2 = - M_P^2$ is the external meson mass of the pseudoscalar meson, $z_k = k \cdot P /|k||P|$, $-1 \leq z_k \leq +1$, is an angle and henceforth Euclidean metric is implicit: 
$k^2 = k_E^2$. The real and imaginary parts of $\mathcal{M}_u(k^2)$ and $\mathcal{F}_u(k^2)$ are plotted in Figs.~\ref{figMReIm} and \ref{figFReIm}, respectively. The behavior of the real part of 
of $\mathcal{M}_u(k^2)$ is smooth, monotonically decreasing in both, real and imaginary $k^2_\pm$ directions, whereas the real part of $\mathcal{F}_u(k^2)$ tends towards its perturbative limit both on
and off the real axis. The imaginary part of both functions is characterized by complex-conjugate extrema near the origin of the parabola. For details of the DSE solutions on the complex plane 
we refer to Refs.~\cite{Fischer:2005en,Krassnigg:2009gd,Rojas:2014aka}. Within the complex momentum range considered herein, the Podolsky model results in complex DSE solutions that differ 
somewhat in magnitude but not much in their analytical behavior from the ones obtained with the gluon model in Ref.~\cite{Qin:2011dd}. However, to really compare the analytic properties of these 
models, a more detailed study on the complex plane is necessary that allows to trace poles and branch cuts outside the parabolic contour presented in Figs.~\ref{figMReIm} and \ref{figFReIm}.


\section{Pseudoscalar Mesons Masses and Decay Constants}
\label{sec:results}

The homogeneous BSE for a $\bar qq$ bound state with relative momentum $p$ and total momentum $P$  can be written as,
\begin{equation}
\label{BSE} 
  \Gamma_{mn}^{fg} (p,P) = \int^\Lambda  \frac{d^{4}k_{E}}{( 2\pi ) ^{4}} \  \mathcal{K}^{kl}_{mn}(p,k,P)  \left  [ S_f (k_+)\Gamma^{fg} (k,P) S_g (k_-) \right ]_{lk} \ ,
\end{equation}
where $m,n,k,l$ collect Dirac and color indices, $f,g$ are flavor indices and $k_+ = k+\eta_+ P, k_- =k- \eta_- P; \eta_+ +\eta_- =1$.  
Since we work within the rainbow-ladder truncation, the BSE kernel  is given by,
\begin{equation}
\label{BSEkernel} 
   \mathcal{K}^{kl}_{mn}(p,k,P) =  - 4\pi \alpha_{\mathrm{eff}}\, \left (\frac{\lambda^a}{2}  \gamma_\mu \right )_{\!\!kn} \!\! P_{\mu\nu}(q) \,
     \left (\frac{\lambda^a}{2}  \gamma_\nu \right )_{\!\! ml } \ ,
\end{equation}
which satisfies the axial-vector WTI~\cite{Maris:1997hd} and as a consequence ensures a massless pion in the chiral limit. As such, Eqs.~\eqref{BSE} and \eqref{BSEkernel} define an 
eigenvalue problem with physical solutions at the on-shell points, $P^2 = -M_P^2$. As in the DSE~\eqref{1}, the vertex renormalization is absorbed in  $\alpha_{\text{eff}}$.

The general Poincar\'e-invariant form of the Bethe-Salpeter amplitudes (BSA), i.e. the solutions of Eq.~(\ref{BSE}), for the  pseudoscalar channel $J^P= 0^-$  in a nonorthogonal base with 
respect to the Dirac trace, $\mathcal{A}^\alpha (p,P) = \gamma_5 \big \{ i\,\mathbb{I}_D$, $\gamma\cdot P$, ${\gamma\cdot p}\, (p\cdot  P)$, $\sigma_{\mu\nu}p_\mu P_\nu \big \}$, 
is given by,
\begin{align}    
  \Gamma  (p, P)  =  \gamma_5  \Big [ \, i\, \mathbb{I}_D  E  (p,P) +  \gamma\cdot P\,  F (p,P)   + \gamma\cdot  p\; p\cdot P\, G  (p,P) + \sigma_{\mu\nu}p_\mu P_\nu\,   H  (p, &  P) \Big ]  \ , 
\label{diracbase}                            
\end{align}
where we suppress color, Dirac and flavor indices for the sake of readability and $\sigma_{\mu\nu} = i/2 [\gamma_\mu,\gamma_\nu ]$. The functions $\mathcal{F}_f ^\alpha (p,P) =$ 
$\big \{ E  (p,P), F  (p,P), G  (p,P), H  (p,P)\big \}$  are Lorentz-invariant scalar amplitudes. For sake of completeness, we note that all BSA are normalized canonically as,
\begin{align}
    2 P_\mu & =  \int^{\Lambda}  \frac{d^{4}k_{E}}{( 2\pi ) ^{4}} \;    \mathrm{Tr}_\mathrm{CD} \left [   \bar \Gamma  (k,-P) \frac{\partial S(k_+)}{\partial P_\mu}
             \Gamma  (k,P) S(k_-)  +  \bar \Gamma  (k,-P) S(k_+) \Gamma  (k,P) \frac{\partial S(k_-)}{\partial P_\mu}  \right ]  \ ,
\label{norm}
\end{align}
where we omit a third term that stems from the derivative of the kernel, $\partial \mathcal{K}^{kl}_{mn}(p,k,P)/ \partial P_\mu$, since it does not contribute in the rainbow-ladder truncation 
of Eq.~\eqref{BSEkernel}\footnote{We verify the values obtained with Eq.~\eqref{norm} with the equivalent normalization condition~\cite{Nakanishi:1965zza,Fischer:2009jm}: 
$(d\ln\lambda_n/dP^2)^{-1}= \mathrm{tr}  \int_k^\Lambda 3 \bar \Gamma(k,-P) S(k_+)\Gamma(k,P)S(k_-)$.}.
In Eq.~(\ref{norm}), the charge-conjugated BSA is defined as $\bar \Gamma (k,-P) := C\, \Gamma^T (-k,-P) C^T$, where $C$ is the charge conjugation operator and the trace
is over Dirac and color indices.  With this normalization we obtain the meson's leptonic decay constants via the integral,
\begin{equation}
  f_P  P_\mu =\int^\Lambda\!  \frac{d^4k_E}{(2\pi)^4} \,\operatorname{Tr}_\mathrm{CD}  \left [ \gamma_5\gamma_\mu\,   S (k_+)\Gamma (k,P) S (k_-)  \right ] \ .
  \label{fdecay} 
\end{equation}
The BSE~\eqref{BSE} is calculated  in Euclidean space and therefore the momenta $k_+$ and $k_-$ of the quark propagators are complex valued. We follow the numerical prescriptions introduced 
in Ref.~\cite{Fischer:2005en}  and refined in Refs.~\cite{Krassnigg:2009gd,Rojas:2014aka} in computing solutions of the DSE in a parabola on the complex plane.


\begin{table}[t!]
\begin{tabular}{ C{1.5cm} | C{2.7cm} C{2.7cm} C{2.3cm}  }                                                                                                \hline\hline
  \phantom{~}                 &  Podolsky Model       &  Q.-C. Model       &     Reference       \\    \hline
 $M_{\pi}$                       & 0.138                        &  0.138      &     0.139~\cite{Tanabashi:2018oca}                                 \\
 $f_{\pi}$                         & 0.133                        &  0.139      &     \multirow{2}{*}{0.130~\cite{Tanabashi:2018oca}}        \\ 
 $f_{\pi}\gmor$                & 0.117                        &   ---            &                                                                                          \\    \hline 
 $M_K$                           & 0.494                        &  0.493      &      0.493~\cite{Tanabashi:2018oca}                                 \\
 $f_K$                             & 0.164                        &  0.164      &     \multirow{2}{*}{0.156~\cite{Tanabashi:2018oca}}         \\  
 $f_K\gmor$                    & 0.162                        &  0.162      &                                                                                           \\    \hline
 $M_{\eta_c(1S)}$           & 2.985                        &  3.065      &      2.984~\cite{Tanabashi:2018oca}                                  \\    
 $f_{\eta_c(1S)}$             & 0.454                        &  0.389      &     \multirow{2}{*}{0.395~\cite{Davies:2010ip}}                   \\
 $f_{\eta_c(1S)}\gmor$    & 0.451                        &  0.380      &                                                                                            \\   \hline
 $M_{D}$                         & 2.100                        &  2.115      &      1.870~\cite{Tanabashi:2018oca}                                   \\    
 $f_{D}$                          & 0.263                         &  0.204      &      0.212~\cite{Aoki:2019cca}                                             \\   \hline
 $M_{D_s}$                    & 2.130                         &  2.130      &      1.968~\cite{Tanabashi:2018oca}                                   \\    
 $f_{D_s}$                      & 0.304                         &  0.249      &      0.250~\cite{Aoki:2019cca}                                             \\    \hline\hline
\end{tabular}  
\caption{Mass spectrum and decay constants for flavor singlet and non-singlet $J^{P}=0^-$ mesons in GeV. We adjusted the current masses to $m_u=m_d=2.64$~MeV,  
$m_s=70$~MeV and  $m_c=865$~MeV in order to reproduce the ground-state masses of the $\pi$, $K$ and $\eta$  mesons, respectively. The weak decay constant
is obtained with Eq.~\eqref{fdecay} and the appropriate GMOR relation. For comparison, we reproduce the values in Ref.~\cite{Rojas:2014aka} using the interaction model 
of Ref.~\cite{Qin:2011dd}.}
\label{table1}
\end{table}


The potential of this effective interaction is illustrated for the pseudoscalar $\bar qq$ channel in Tab.~\ref{table1}. We work in the isospin-symmetric limit $m_u = m_d$ and set the light quark's 
mass scale at 19~GeV with the pion mass; analogously we fix the strange- and charm-quark masses with the kaon and the $\eta_c$. The resulting weak decay constant of the pion is within 
2\% of the experimentally extracted value and that of the kaon is within 5\% of its reference value. The charmonium's decay constant is about 15\% larger than a calculation using lattice QCD. 
We also compare the weak decay constants obtained with Eq.~\eqref{fdecay} with those making use of the Gell-Mann-Oaks-Renner (GMOR) relation described in detail, for example,
in Ref.~\cite{Rojas:2014aka} and find very good agreement. 

This is a first step to establish the model's potential to correctly describe light-meson and quarkonia ground states and as additional check we calculate the $D$ and $D_s$ meson's masses 
and decay constants. As observed in Tab.~\ref{table1} and previously in Ref.~\cite{Rojas:2014aka}, their masses are overestimated by 12\% for the $D$ and 8\% for the $D_s$ with respect to 
experimental values. This is a consequence of the rainbow-ladder truncation which neglects the dramatically different impact of vertex dressing for heavy and light quarks and can be strongly 
improved by inclusion of this effect~\cite{Serna:2020txe}. Likewise, the weak decay constants are also  22\% larger than results reported by the FLAG collaboration~\cite{Aoki:2019cca}. 
Nonetheless, the Podolsky propagator facilitates the numerical treatment of the DSE with a large external Euclidean mass (using $\eta_+ = 0.8$ and $\eta_-=0.2$)  on the complex plane in 
comparison with the model of Ref.~\cite{Qin:2011dd} and both the iterative treatments of the DSE  and BSE mesons converge rapidly in case of the $D$ and $D_s$.


\section{Summary and Conclusions}
\label{conclude}

We solve for the first time the DSE with the Podolsky propagator in rainbow truncation and find a mass gap for a typical hadronic scale of the Podolsky mass. Based on this observation, we
propose a novel interaction model within the rainbow-ladder truncation of the DSE and BSE kernels, based on the Podolsky propagator which preserves gauge invariance in perturbative 
GQED. However, in our approach we merely interpret this massive propagator as a nonperturbative ansatz for the dressed gluon propagator. The associated mass scale we find is reminiscent of earlier 
DSE studies of the gluon. The Podolsky propagator has the same integrability properties as the perturbative propagator, a feature which is the practical motivation for our model, and when employed 
in the appropriate DSE we find well behaved solutions on the complex plane even for larger time-like momenta. Employing these complex  solutions for the quark propagators in the BSE, we fix the 
quark masses and interaction parameters with the masses of the $\pi$, $K$, $\eta_c$ and find weak decay constants that agree very well with experimental reference values or results from lattice-QCD 
simulations. The $D$ mesons are also obtained within this framework, where their masses are somewhat overestimated and the mass difference between the $D$ and $D_s$ is too small, a 
consequence of the too simplistic truncation for heavy-light systems.

The Podolsky propagator along with the ansatz in Eq.~\eqref{eq:g2} ought to be used in future calculations of radial excitations and other $J^{PC}$ channels. Since we are interested 
in static properties of the mesons, we here omit the perturbative term of the interaction commonly included in other models. As a consequence of this simple form, we established that the angular 
integration of the DSE, Eqs.~\eqref{Eq.11} and \eqref{Eq.12}, can be performed analytically. Likewise, the simple pole structure of the Podolsky propagator allows, in given cases,  for analytical 
calculations as with any perturbative propagator. These features are attractive enough to raise one's curiosity about possible solutions of the DSE and BSE in Minkowski space, which is of practical
importance with regard to the study of parton distribution functions defined on the light front. Amongst other applications, one may explore this interaction to obtain more sophisticated quark propagators 
and wave functions at finite density to understand the pion's properties in a dense nuclear medium~\cite{deMelo:2014gea}.


\acknowledgments 
We acknowledge financial support from ``Patrimonio Aut\'onomo Fondo Nacional de Financiamiento para la Ciencia, la Tecnolog\'ia y la Innovaci\'on, Francisco Jos\'e de Caldas'', 
from  Funda\c{c}\~ao de Amparo \`a Pesquisa do Estado de S\~ao Paulo, grant no.~2018/20218-4, and Conselho Nacional de Desenvolvimento Cient\'ifico e Tecnol\'ogico, 
grant no.~428003/2018-4. This work was also partly supported by the ``Vicerrector\'ia de Investigaciones e Interacci\'on Social VIIS de la Universidad de Nari\~no,  project numbers 
1928 and 2172.  B.E. appreciated the hospitality of Universidade de Nari\~no during his stay in San Juan de Pasto and participates in the INCT-FNA project no.~464898/2014-5.
Insightful comments on the manuscript by Fernando Serna were strongly appreciated.


\appendix

\section{Dyson-Schwinger Equation with the Podolsky Propagator}
\label{sec:appendixA}

For a a quark propagator, the DSE in Minkowski space is described by a nonlinear integral equation,
\begin{equation}
S^{-1} ( p ) =  Z_2\gamma\cdot  p  -  Z_4 m(\mu)  +  iZ_1g^{2}C_F  \int \! \frac{d^{4}k}{( 2\pi )^{4}}\,  \gamma _{\mu }S ( k ) \,
                                    \Gamma_{\nu }\left( k,p\right) G^{\mu \nu } ( q ) \  ,
                                    \label{Eq.1}
\end{equation}
with $C_F =4/3$ in the fundamental representation of SU(3). 
The GQED gauge propagator in a covariant gauge, specified by the gauge parameter $\xi $, is given by,
\begin{align}
  G_{\mu\nu}= \Delta(q^2) P_{\mu \nu }(q),
\end{align}
with $q=k-p$ and where,
\begin{eqnarray}
  P_{\mu \nu }( q)  =    \Delta_{\mu \nu }( q) -\left[ g_{\mu \nu }\smallskip +\left( 1-\xi \right) \frac{q_{\mu }q_{\nu }}{q^{2}-m_{P}^{2}}  \right ] 
    \frac{1}{\smallskip \smallskip q^{2}-m_{P}^{2}} +  \left( \smallskip 1-2\xi \right) \frac{q_{\mu }q_{\nu }}{q^{2}  \left  ( q^{2}-m_{P}^{2}\right) } 
   +   \frac{q_{\mu }q_{\nu }}{\left( q^{2}-m_{P}^{2}\right) ^{2}} \ .
   \label{Eq.2}
\end{eqnarray}
Here, $\Delta_{\mu \nu }( q) $ is the contribution of the Maxwell theory:
\begin{equation}
     \Delta_{\mu \nu }( q) =\frac{1 }{q^{2}}  \left( g_{\mu \nu }-\frac{q_{\mu }q_{\nu }}{q^{2}}\right) + \xi\, \frac{q_{\mu }q_{\nu }}{q^4 } \ .
  \label{Eq.3}
\end{equation}

We employ for the vertex structure its bare form, which is the rainbow truncation:
\begin{equation}
  \Gamma _{\mu }   \longrightarrow \  \gamma _{\mu } \ .  
\label{Eq.5}
\end{equation}%
In general, the Dirac structure of the fermion propagator depends on two independent functions, the wave function renormalization $\mathcal{F} ( p )$ and the mass function 
$ \mathcal{M}_f ( p ) $, such that:
\begin{equation}
  S ( p ) =  \frac{\mathcal{F} ( p ) }{\gamma \cdot  p - \mathcal{M} (p ) }\  .  
  \label{Eq.6}
\end{equation}
With this, expression~\eqref{Eq.1} can be rewritten as:
\begin{equation}
    \frac{\gamma \cdot p - \mathcal{M} ( p ) }{\mathcal{F} ( p ) }  =  Z_2\gamma \cdot p  - Z_4   m(\mu) 
    +  iZ_1 g^{2} C_F\int \frac{d^{4}k}{(2\pi )^4} \, \mathcal{F}(k) \gamma_\mu\, \frac{ \gamma  \cdot k + \mathcal{M} (k) }{k^2 - \mathcal{M}^{2} (k)}\, G^{\mu \nu} (q) \,\gamma _\nu \ .
\label{Eq.7}
\end{equation}%
Taking the trace of Eq.~\eqref{Eq.7} results in the expression,
\begin{equation}
     \frac{ \mathcal{M} (p) }{\mathcal{F} ( p ) } =  Z_4   m(\mu)  - i Z_1 g^{2} C_F\int \frac{d^{4}k}{( 2\pi ) ^{4}}\frac{ \mathcal{M} ( k ) \mathcal{F}  (k ) }{k^2 - \mathcal{M}^2 (k) }
     \, G_{\ \mu }^{\mu } ( q ) \ ,  
 \label{Eq.8}
\end{equation}
where $G^{\mu}_{\ \mu}(q) = \Delta(q^2)P^{\mu}_{\ \mu}(q)$ and,
\begin{eqnarray}
      P_{\  \mu }^{\mu } ( q ) & = &  \Delta^{\mu }_{\ \mu} ( q ) - \left[ g_{\ \mu}^{\mu}  + ( 1 - \xi )\, \frac{q^{\mu }q_{\mu} }{q^2-m_P^2 }  \right ] 
      \frac{1}{q^2-m_P^2}  +  ( 1-2\xi) \frac{q^{\mu }q_{\mu } } {q^2 ( q^2 -m_{P}^2 ) }
     + \frac{q^{\mu }q_{\mu }}{\left( q^{2}-m_{P}^{2}\right) ^{2}}   \nonumber \\
                                       & = &  \Delta _{\ \mu }^{\mu} ( q ) -  ( 3+2\xi ) \frac{1}{ q^2 -m_P^2  } +  \xi\, \frac{q^2}{ ( q^2-m_P^2 )^2 } \ ,
\label{Eq.9}
\end{eqnarray}
with,
\begin{equation*}
   \Delta _{~\mu }^{\mu } ( q )  =  \frac{1}{q^2} \left( g_{~\mu }^{\mu } - \frac{q^{\mu }q_{\mu }}{q^{2}}\right) +  \xi \frac{q^{\mu }q_{\mu }}{q^4 }
                                               =   ( 3+\xi ) \frac{1}{q^2 } \ .
\end{equation*}
It follows that,
\begin{eqnarray}
   \frac{\mathcal{M} (p)}{\mathcal{F} (p) } &=&   Z_4 m(\mu) - iZ_1g^{2}C_F\int \frac{d^{4}k}{( 2\pi ) ^{4}}\frac{ \mathcal{M} ( k ) \mathcal{F} ( k) }{k^2- \mathcal{M}^2 (k)} \Delta(q^2)
        \left[ \Delta _{~\mu }^{\mu }( q) - ( 3+2\xi ) \frac{1}{ q^2-m_P^2  } + \xi \frac{q^2}{ \left( q^2 -m_P^2 \right )^2 } \right ]  \notag \\
                                                                &= & \left. \frac{ \mathcal{M} ( p ) }{\mathcal{F} (p) }\right\vert_M  + \left. \frac{\mathcal{M }( p) }{\mathcal{F}(p) }\right\vert_P \ ,  
 \label{Eq.10}
\end{eqnarray}
where the two terms are, 
\begin{eqnarray}
  \left. \frac{ \mathcal{M}( p ) }{\mathcal{F} ( p) } \right\vert _{M} &\equiv & 
  Z_4 m(\mu) - i Z_1g^{2}C_F\int \frac{d^{4}k}{( 2\pi )^{4}}\frac{ \mathcal{M}( k) \mathcal{F} ( k) }{k^2 - \mathcal{M}^2( k) }\Delta(q^2)\Delta_{~\mu }^{\mu }( q)  
\notag \\
            & = &  Z_4 m(\mu)-  iZ_1g^{2}C_F ( 3+\xi ) \int\! \frac{d^{4}k}{( 2\pi )^{4}}\frac{ \mathcal{M}( k)  \mathcal{F}( k)}{k^{2} -  \mathcal{M}^{2}( k) }\frac{\Delta(q^2)}{q^2}, 
 \label{Eq.11}
\end{eqnarray}%
which is the Maxwell contribution and, 
\begin{equation}
    \left. \frac{ \mathcal{M} ( p)}{\mathcal{F}( p) }\right\vert_{P}  \equiv ( 3+2\xi ) iZ_1g^{2}C_F \int \frac{d^{4}k}{( 2\pi ) ^4}
            \frac{ \mathcal{M}( k) \mathcal{F} ( k) }{k^{2}- \mathcal{M}^{2}( k) }\frac{\Delta(q^2)}{q^{2}-m_{P}^{2}}
            - iZ_1g^{2}  C_F  \xi \int \frac{d^{4}k}{( 2\pi ) ^{4}}\frac{ \mathcal{M}( k)  \mathcal{F}( k)}{k^{2}- \mathcal{M}^{2}( k) }\frac{q^{2}\Delta(q^2)}{\left( q^2-m_P^{2}\right) ^2} \, ,
\label{Eq.12}
\end{equation}%
is due to Podolsky's GQED extension.

Now, if we multiply Eq.~\eqref{Eq.7} by  $\gamma ^{\beta }p_{\beta }$ and take the trace, we can project out the wave function renormalization $\mathcal{F} ( p )$, such that, 
\begin{equation}
   \frac{p^{2}}{\mathcal{F}( p)}=Z_2 p^{2}+iZ_1g^2C_F\int \! \frac{d^{4}k}{( 2\pi ) ^{4}} \, \mathcal{F}( k)\,    \frac{  2p_{\mu}k_{\nu }P^{\mu \nu } ( q ) 
       - p \cdot k \, P_{~\mu }^{\mu } ( q )  }{k^{2}- \mathcal{M}^2( k) }  \, \Delta(q^2)\ ,
\label{Eq.13}
\end{equation}%
where the expression in the numerator is found to be:
\begin{eqnarray*}
     2p_{\mu }k_{\nu }P^{\mu \nu }( q) - p\cdot k  \, P_{~\mu }^{\mu }( q) 
     & = &  \left[ \left( \xi -3\right) \frac{\left( p\cdot k\right) }{q^{2}}+2\left( \xi -1\right) \frac{1}{q^{4}}\left[ \left( p\cdot k\right) ^{2}-p^{2}k^{2}\right] \right]  \\
     & + &  \left[ ( 3-2\xi ) ( p\cdot k ) +2 ( 1-2\xi ) \, \frac{ (p\cdot k  )^{2}-p^{2}k^{2}  }{q^{2}} \right  ]  \frac{1}{ q^2 - m_P^2  }   \\
     & + & \Big [ \, \xi \left( p\cdot k\right) q^{2}+2\xi \left( ( p\cdot k )^2  -p^2 k^2  \right) \Big ] \frac{1}{ \left( q^{2}-m_P^2 \right )^{2}} \ .
\end{eqnarray*}
We can again separate the integral into a Maxwell and Podolsky contribution, 
\begin{equation}
   \frac{1}{\mathcal{F}( p)}=\left. \frac{1}{\mathcal{F} ( p ) }\right\vert _{M}+\left. \frac{1}{\mathcal{F}( p)} \right\vert _{P}\ , 
 \label{Eq.14}
\end{equation}
where,
\begin{equation}
   \left. \frac{1}{\mathcal{F}( p)} \right \vert _{M}  \equiv \,  Z_2  +  \frac{ i Z_1g^2C_F}{p^{2}}
   \int \! \frac{d^{4}k} {( 2\pi )^{4}}  \frac{\mathcal{F} ( k) \Delta(q^2) }{k^{2}- \mathcal{M}^{2}( k) }
   \left[ ( \xi -3 ) \frac{ p\cdot k }{q^2} + 2 ( \xi -1 ) \frac{1}{q^{4}}\left[ ( p\cdot k ) ^{2}-p^{2}k^{2}\right]  \right  ] ,
\label{Eq.15}
\end{equation}
and
\begin{eqnarray}
   \left. \frac{1}{\mathcal{F}( p)}\right\vert _{P} &\equiv &\frac{ iZ_1 g^2C_F}{p^2}  \int \frac{d^{4}k}{( 2\pi )^{4}}\frac{\mathcal{F} ( k)\Delta(q^2) }{k^{2}- \mathcal{M}^{2}( k) }
   \left[ ( 3-2\xi  )\,  p\cdot k  +2 (1-2\xi ) \frac{  ( p\cdot k )^2 -p^2 k^2  }{q^{2}} \right ] \frac{1}{ q^{2}-m_{P}^{2}  }  \notag \\
  & + &  \frac{iZ_1 g^2C_F}{p^{2}}\xi \int \frac{d^{4}k}{( 2\pi)^4 } \frac{ \mathcal{F}( k)\Delta(q^2) }{k^{2}- \mathcal{M}^{2}( k) }
             \Big [ p\cdot k\, q^2 + 2 \left \{  ( p\cdot k )^2 -p^2 k^2  \right \}  \Big ] \frac{1}{ (q^2 -m_P^2  )^2  } \ .  
\label{Eq.16}
\end{eqnarray}


\subsection{Landau Gauge}

In Landau gauge, Eq.~\eqref{Eq.10} takes the form,
\begin{eqnarray}
    \left. \frac{ \mathcal{M}( p)}{\mathcal{F}( p)} \right\vert _{M}    & = &   Z_4 m(\mu)-3iZ_1 g^2C_F
          \!  \int \frac{d^{4}k}{ ( 2\pi)^{4}}  \, \frac{ \mathcal{M}( k)  \mathcal{F}( k)}{k^{2}-\mathcal{M}^{2}( k) }\frac{\Delta(q^2)}{q^{2}}, \\
   \left. \frac{ \mathcal{M} ( p)}{\mathcal{F}( p)}  \right\vert _{P}    &  =  &     3iZ_1 g^2 C_F
        \!   \int \frac{d^{4}k}{( 2\pi )^{4}}  \, \frac{ \mathcal{M}( k)  \mathcal{F}( k)}{k^{2}- \mathcal{M}^{2}( k) }\frac{\Delta(q^2)}{q^ 2-m_{P}^2}  \ .
\end{eqnarray}
If we apply a  Wick rotation we obtain the Euclidean space expressions for these integrals,
\begin{eqnarray}
    \left. \frac{ \mathcal{M} \left( p_{E}\right) }{\mathcal{F}\left( p_{E}\right) }\right\vert _{M} &  = &   Z_4m(\mu) + 3Z_1 g^2C_F
       \int \! \frac{d^{4}k_{E}}{( 2\pi )^4} \,  \frac{ \mathcal{M}(k_{E} ) \mathcal{F} (k_{E}) }{k_{E}^{2}+ \mathcal{M}^2 ( k_{E} ) }\frac{\Delta(q^2_E)}{q_{E}^2} \ , \\
    \left. \frac{ \mathcal{M} \left( p_{E}\right) }{\mathcal{F}\left( p_{E}\right) }\right\vert _{P} &=&-\smallskip 3Z_1 g^2C_F
          \int \frac{d^{4}k_{E}}{ ( 2\pi)^4}  \,  \frac{ \mathcal{M}(k_{E} ) \mathcal{F} ( k_{E} ) }{k_{E}^{2}+ \mathcal{M}^{2}( k_{E} ) }  \frac{\Delta(q^2_E)}{q_{E}^2+m_{P}^2} \ .
\end{eqnarray}
and we obtain explicitly, 
\begin{equation}
  \frac{ \mathcal{M}_f\left( p_{E}\right) }{\mathcal{F}\left( p_{E}\right) } = Z_4m(\mu) + 3Z_1 g^2C_F  
    \int \frac{d^{4}k_{E}}{( 2\pi )^{4}} \, \frac{\mathcal{M}( k_{E} ) \mathcal{F}( k_{E} )  }{k_{E}^2+ \mathcal{M}^{2}( k_{E}) }\left( \frac{1}{q_{E}^2}-\frac{1}{q_{E}^2+m_P^2} \right ) \Delta(q^2_E)\ .  
\label{MLandau}
\end{equation}
Similarly, Eq.~\eqref{Eq.14} becomes in Landau gauge,
\begin{eqnarray}
      \left. \frac{1}{\mathcal{F}( p)}\right\vert _{M}   &  =  &  Z_2-\frac{ iZ_1 g^2C_F}{p^{2}} 
          \int \!\frac{d^{4}k}{( 2\pi )^4}  \,  \frac{\mathcal{F}( k) }{k^{2}- \mathcal{M}^{2}( k) }\left[ 3 p\cdot k  +2\, \frac{  \left( p\cdot k\right) ^{2}-p^{2}k^{2}  }{q^2 }\right] 
          \frac{\Delta(q^2)}{q^{2}} \ , \\
     \left. \frac{1}{\mathcal{F}( p)}\right\vert _{P}    &  =  &  \frac{iZ_1 g^2C_F}{p^{2}}
          \int \! \frac{d^{4}k}{( 2\pi )^4}  \, \frac{\mathcal{F} ( k) }{k^{2}- \mathcal{M}^{2}( k) }\left[ 3  p\cdot k  +2\, \frac{ \left( p\cdot k\right) ^{2}-p^{2}k^{2}  }{q^{2} } \right ] 
          \frac{\Delta(q^2)} { q^2-m_P^2  } \ .
\end{eqnarray}
In Euclidean space these integrals are given by, 
\begin{eqnarray}
  \left. \frac{1}{\mathcal{F} ( p_{E} ) } \right \vert _{M}   & = &   Z_2 + \frac{ Z_1 g^2C_F}{p_{E}^{2} }
         \int  \frac{d^{4}k_{E}}{( 2\pi )^4} \, \frac{\mathcal{F}( k_{E}) }{k_{E}^{2} + \mathcal{M}^{2}( k_{E})  } 
          \left[ 3 p_{E}\cdot k_{E}  + 2\, \frac{\left ( p_{E}\cdot k_{E}\right )^2  - p_{E}^{2} k_{E}^{2}  }{q_E^2} \right ]  \frac{\Delta(q^2_E)}{q_{E}^2} \ , \\
   \left. \frac{1}{\mathcal{F} ( p_{E} ) }\right\vert _{P} &  =  &  - \frac{Z_1 g^2C_F}{p_{E}^{2} }
   \int \frac{d^{4}k_{E}}{( 2\pi)^4} \,  \frac{\mathcal{F}( k_{E})}{k_{E}^{2}+ \mathcal{M}^{2} ( k_{E}) }\left[ 3 p_{E}\cdot k_{E} 
   +2\, \frac{ \left( p_{E}\cdot k_{E}\right) ^{2}-p_{E}^{2}k_{E}^{2}  }{q_{E}^{2}}\right] \frac{\Delta(q^2_E)}{q_E^2+m_P^2  } \   .
\end{eqnarray}
Adding the two contributions we finally arrive at the Euclidean-space integral equation,
\begin{equation}
    \frac{1}{\mathcal{F}\left( p_{E}\right) }   =  Z_2 +  \frac{Z_1 g^2C_F}{p_{E}^{2}}\int \frac{d^{4}k_{E}}{( 2\pi)^{4}}\frac{\mathcal{F} (k_{E}) }{k_{E}^{2}+ \mathcal{M}^{2}( k_{E} ) }
    \left[ 3  p_{E}\cdot k_{E} + 2\, \frac{  \left( p_{E}\cdot k_{E} \right)^{2}-p_{E}^{2}k_{E}^{2}  }{q_{E}^{2}}\right] \left[ \frac{1}{q_{E}^2} -\frac{1}{ q_E^2+m_{P}^2 }  \right ]  \Delta(q^2_E) \ .
\label{Flandau}
\end{equation}
This simplifies to, 
\begin{equation}
   \frac{1}{\mathcal{F}\left( p_{E}\right) }   =  Z_2 +  \frac{Z_1 g^2C_F}{p_{E}^{2}}\int \frac{d^{4}k_{E}}{( 2\pi)^{4}}\frac{\mathcal{F} (k_{E}) }{k_{E}^{2}+ \mathcal{M}^{2}( k_{E} ) }
    \left[ 3  p_{E}\cdot k_{E} + 2\, \frac{  \left( p_{E}\cdot k_{E} \right)^{2}-p_{E}^{2}k_{E}^{2}  }{q_{E}^{2}}\right] \frac{m_P^2}{ q_E^2+m_{P}^2 }   \frac{\Delta(q^2_E)}{q_E^2} \ ,
\end{equation}
which justifies the definition of $\mathcal{G}(q_E^2)$ in Eq.~\eqref{eq:g2} in the particular case of Landau gauge and $\Delta(q^2_E)= 1$. 
It turns out that, the angular integration of Eqs.~\eqref{MLandau} and \eqref{Flandau} can be performed analytically using, 
\begin{eqnarray*}
   p_{E}\cdot k_{E}  & = &p_{E}k_{E}\cos \theta _{1}, \\ 
     \left( p_{E}\cdot k_{E}\right) ^{2}-p_{E}^{2}k_{E}^{2}   &=&-p_{E}^{2}k_{E}^{2}\sin ^{2}\theta _{1} \ ,
\end{eqnarray*}
and working in the rest frame where $p_{E}^{\mu }=(p_{E},0,0,0)$.



\begin{thebibliography}{99}

\bibitem{Mandelstam:1979xd} 
  S.~Mandelstam,
  Phys.\ Rev.\ D {\bf 20}, 3223 (1979).
  doi:10.1103/PhysRevD.20.3223

\bibitem{Cornwall:1974vz}
J.~M.~Cornwall, R.~Jackiw and E.~Tomboulis,
Phys. Rev. D \textbf{10}, 2428-2445 (1974)
doi:10.1103/PhysRevD.10.2428

\bibitem{Cornwall:1981zr}
J.~M.~Cornwall,
Phys. Rev. D \textbf{26}, 1453 (1982)
doi:10.1103/PhysRevD.26.1453

\bibitem{Cornwall:1989gv}
J.~M.~Cornwall and J.~Papavassiliou,
Phys. Rev. D \textbf{40}, 3474 (1989)
doi:10.1103/PhysRevD.40.3474

\bibitem{Burden:1993gy}
C.~J.~Burden and C.~D.~Roberts,
Phys. Rev. D \textbf{47}, 5581-5588 (1993)
doi:10.1103/PhysRevD.47.5581
[arXiv:hep-th/9303098 [hep-th]].

\bibitem{Brown:1988bn} 
  N.~Brown and M.~R.~Pennington,
  Phys.\ Rev.\ D {\bf 39}, 2723 (1989).
  doi:10.1103/PhysRevD.39.2723
  
\bibitem{Fischer:2004nq}
C.~S.~Fischer, R.~Alkofer, T.~Dahm and P.~Maris,
Phys. Rev. D \textbf{70}, 073007 (2004)
doi:10.1103/PhysRevD.70.073007
[arXiv:hep-ph/0407104 [hep-ph]].

\bibitem{Bashir:2005wt}
A.~Bashir and A.~Raya,
Few Body Syst. \textbf{41}, 185-199 (2007)
doi:10.1007/s00601-007-0177-3
[arXiv:hep-ph/0511291 [hep-ph]].

\bibitem{Chang:2009ae}
L.~Chang, I.~C.~Clo\"et, B.~El-Bennich, T.~Kl\"ahn and C.~D.~Roberts,
Chin. Phys. C \textbf{33}, 1189-1196 (2009)
doi:10.1088/1674-1137/33/12/022
[arXiv:0906.4304 [nucl-th]].

\bibitem{Bashir:2013zha}
A.~Bashir, A.~Raya and J.~Rodr\'iguez-Quintero,
Phys. Rev. D \textbf{88}, 054003 (2013)
doi:10.1103/PhysRevD.88.054003
[arXiv:1302.5829 [hep-ph]].
  
\bibitem{Bashir:2012fs}
A.~Bashir, L.~Chang, I.~C.~Clo\"et, B.~El-Bennich, Y.~X.~Liu, C.~D.~Roberts and P.~C.~Tandy,
Commun. Theor. Phys. \textbf{58}, 79-134 (2012)
doi:10.1088/0253-6102/58/1/16
[arXiv:1201.3366 [nucl-th]].

\bibitem{Cloet:2013jya}
I.~C.~Clo\"et and C.~D.~Roberts,
Prog. Part. Nucl. Phys. \textbf{77}, 1-69 (2014)
doi:10.1016/j.ppnp.2014.02.001
[arXiv:1310.2651 [nucl-th]].

\bibitem{Munczek:1983dx} 
  H.~J.~Munczek and A.~M.~Nemirovsky,
  Phys.\ Rev.\ D {\bf 28}, 181 (1983).
  doi:10.1103/PhysRevD.28.181
  
\bibitem{Tissier:2010ts}
M.~Tissier and N.~Wschebor,
Phys. Rev. D \textbf{82} (2010), 101701
doi:10.1103/PhysRevD.82.101701
[arXiv:1004.1607 [hep-ph]].

\bibitem{Pelaez:2017bhh}
M.~Pel\'aez, U.~Reinosa, J.~Serreau, M.~Tissier and N.~Wschebor,
Phys. Rev. D \textbf{96}, no.11, 114011 (2017)
doi:10.1103/PhysRevD.96.114011
[arXiv:1703.10288 [hep-th]].

\bibitem{Munczek:1988er} 
  H.~J.~Munczek and D.~W.~McKay,
  Phys.\ Rev.\ D {\bf 39}, 888 (1989)
  Erratum: [Phys.\ Rev.\ D {\bf 46}, 5209 (1992)].
  doi:10.1103/PhysRevD.46.5209, 10.1103/PhysRevD.39.888

\bibitem{Praschifka:1989fd} 
  J.~Praschifka, R.~T.~Cahill and C.~D.~Roberts,
  Int.\ J.\ Mod.\ Phys.\ A {\bf 4}, 4929 (1989).
  doi:10.1142/S0217751X89002090

\bibitem{Williams:1989tv} 
  A.~G.~Williams, G.~Krein and C.~D.~Roberts,
  Annals Phys.\  {\bf 210}, 464 (1991).
  doi:10.1016/0003-4916(91)90051-9

\bibitem{vonSmekal:1991fp} 
  L.~von Smekal, P.~A.~Amundsen and R.~Alkofer,
  Nucl.\ Phys.\ A {\bf 529}, 633 (1991).
  doi:10.1016/0375-9474(91)90589-X

\bibitem{Jain:1993qh} 
  P.~Jain and H.~J.~Munczek,
  Phys.\ Rev.\ D {\bf 48}, 5403 (1993)
  doi:10.1103/PhysRevD.48.5403
  [hep-ph/9307221].

\bibitem{Frank:1995uk} 
  M.~R.~Frank and C.~D.~Roberts,
  Phys.\ Rev.\ C {\bf 53}, 390 (1996)
  doi:10.1103/PhysRevC.53.390
  [hep-ph/9508225].

\bibitem{Maris:1997tm} 
  P.~Maris and C.~D.~Roberts,
  Phys.\ Rev.\ C {\bf 56}, 3369 (1997)
  doi:10.1103/PhysRevC.56.3369
  [nucl-th/9708029].
  
\bibitem{Maris:1997hd} 
  P.~Maris, C.~D.~Roberts and P.~C.~Tandy,
  Phys.\ Lett.\ B {\bf 420}, 267 (1998)
  doi:10.1016/S0370-2693(97)01535-9
  [nucl-th/9707003].

\bibitem{Maris:1999nt} 
  P.~Maris and P.~C.~Tandy,
  Phys.\ Rev.\ C {\bf 60}, 055214 (1999)
  doi:10.1103/PhysRevC.60.055214
  [nucl-th/9905056].

\bibitem{Alkofer:2002bp}
R.~Alkofer, P.~Watson and H.~Weigel,
Phys. Rev. D \textbf{65} (2002), 094026
doi:10.1103/PhysRevD.65.094026
[arXiv:hep-ph/0202053 [hep-ph]].

\bibitem{Qin:2011dd} 
  S.~x.~Qin, L.~Chang, Y.~x.~Liu, C.~D.~Roberts and D.~J.~Wilson,
  Phys.\ Rev.\ C {\bf 84}, 042202 (2011)
  doi:10.1103/PhysRevC.84.042202
  [arXiv:1108.0603 [nucl-th]].

\bibitem{Chang:2011ei}
L.~Chang and C.~D.~Roberts,
Phys. Rev. C \textbf{85}, 052201 (2012)
doi:10.1103/PhysRevC.85.052201
[arXiv:1104.4821 [nucl-th]].

\bibitem{Chang:2013pq}
L.~Chang, I.~C.~Clo\"et, J.~J.~Cobos-Mart\'inez, C.~D.~Roberts, S.~M.~Schmidt and P.~C.~Tandy,
Phys. Rev. Lett. \textbf{110}, no.13, 132001 (2013)
doi:10.1103/PhysRevLett.110.132001
[arXiv:1301.0324 [nucl-th]].

\bibitem{Chang:2013nia}
L.~Chang, I.~C.~Clo\"et, C.~D.~Roberts, S.~M.~Schmidt and P.~C.~Tandy,
Phys. Rev. Lett. \textbf{111}, no.14, 141802 (2013)
doi:10.1103/PhysRevLett.111.141802
[arXiv:1307.0026 [nucl-th]].

\bibitem{Rojas:2014aka}
E.~Rojas, B.~El-Bennich and J.~P.~B.~C.~de Melo,
Phys. Rev. D \textbf{90}, 074025 (2014)
doi:10.1103/PhysRevD.90.074025
[arXiv:1407.3598 [nucl-th]].

\bibitem{Raya:2015gva}
K.~Raya, L.~Chang, A.~Bashir, J.~J.~Cobos-Martinez, L.~X.~Guti\'errez-Guerrero, C.~D.~Roberts and P.~C.~Tandy,
Phys. Rev. D \textbf{93}, no.7, 074017 (2016)
doi:10.1103/PhysRevD.93.074017
[arXiv:1510.02799 [nucl-th]].

\bibitem{El-Bennich:2016qmb}
B.~El-Bennich, G.~Krein, E.~Rojas and F.~E.~Serna,
Few Body Syst. \textbf{57}, no.10, 955-963 (2016)
doi:10.1007/s00601-016-1133-x
[arXiv:1602.06761 [nucl-th]].

\bibitem{El-Bennich:2016bno}
B.~El-Bennich, M.~A.~Paracha, C.~D.~Roberts and E.~Rojas,
Phys. Rev. D \textbf{95}, no.3, 034037 (2017)
doi:10.1103/PhysRevD.95.034037
[arXiv:1604.01861 [nucl-th]].

\bibitem{Mojica:2017tvh}
F.~F.~Mojica, C.~E.~Vera, E.~Rojas and B.~El-Bennich,
Phys. Rev. D \textbf{96}, no.1, 014012 (2017)
doi:10.1103/PhysRevD.96.014012
[arXiv:1704.08593 [hep-ph]].

\bibitem{Shi:2018zqd}
C.~Shi and I.~C.~Clo\"et,
Phys. Rev. Lett. \textbf{122}, no.8, 082301 (2019)
doi:10.1103/PhysRevLett.122.082301
[arXiv:1806.04799 [nucl-th]].

\bibitem{Serna:2020txe}
F.~E.~Serna, R.~Correa da Silveira, J.~J.~Cobos-Mart\'\i{}nez, B.~El-Bennich and E.~Rojas,
[arXiv:2008.09619 [hep-ph]].

\bibitem{Qin:2020jig}
S.~x.~Qin and C.~D.~Roberts,
[arXiv:2009.13637 [hep-ph]].

\bibitem{Cloet:2008re}
I.~C.~Clo\"et, G.~Eichmann, B.~El-Bennich, T.~Klahn and C.~D.~Roberts,
Few Body Syst. \textbf{46}, 1-36 (2009)
doi:10.1007/s00601-009-0015-x
[arXiv:0812.0416 [nucl-th]].

\bibitem{Eichmann:2009qa}
G.~Eichmann, R.~Alkofer, A.~Krassnigg and D.~Nicmorus,
Phys. Rev. Lett. \textbf{104}, 201601 (2010)
doi:10.1103/PhysRevLett.104.201601
[arXiv:0912.2246 [hep-ph]].

\bibitem{Aznauryan:2012ba}
I.~G.~Aznauryan, A.~Bashir, V.~Braun, S.~J.~Brodsky, V.~D.~Burkert, L.~Chang, C.~Chen, B.~El-Bennich, I.~C.~Clo\"et and P.~L.~Cole, \textit{et al.}
Int. J. Mod. Phys. E \textbf{22}, 1330015 (2013)
doi:10.1142/S0218301313300154
[arXiv:1212.4891 [nucl-th]].

\bibitem{Segovia:2015hra}
J.~Segovia, B.~El-Bennich, E.~Rojas, I.~C.~Clo\"et, C.~D.~Roberts, S.~S.~Xu and H.~S.~Zong,
Phys. Rev. Lett. \textbf{115}, no.17, 171801 (2015)
doi:10.1103/PhysRevLett.115.171801
[arXiv:1504.04386 [nucl-th]].

\bibitem{Eichmann:2016hgl}
G.~Eichmann, C.~S.~Fischer and H.~Sanchis-Alepuz,
Phys. Rev. D \textbf{94}, no.9, 094033 (2016)
doi:10.1103/PhysRevD.94.094033
[arXiv:1607.05748 [hep-ph]].

\bibitem{Eichmann:2016yit}
G.~Eichmann, H.~Sanchis-Alepuz, R.~Williams, R.~Alkofer and C.~S.~Fischer,
Prog. Part. Nucl. Phys. \textbf{91}, 1-100 (2016)
doi:10.1016/j.ppnp.2016.07.001
[arXiv:1606.09602 [hep-ph]].

\bibitem{Sanchis-Alepuz:2017jjd}
H.~Sanchis-Alepuz and R.~Williams,
Comput. Phys. Commun. \textbf{232}, 1-21 (2018)
doi:10.1016/j.cpc.2018.05.020
[arXiv:1710.04903 [hep-ph]].

\bibitem{Chen:2017pse}
C.~Chen, B.~El-Bennich, C.~D.~Roberts, S.~M.~Schmidt, J.~Segovia and S.~Wan,
Phys. Rev. D \textbf{97}, no.3, 034016 (2018)
doi:10.1103/PhysRevD.97.034016
[arXiv:1711.03142 [nucl-th]].

\bibitem{Chen:2018nsg}
C.~Chen, Y.~Lu, D.~Binosi, C.~D.~Roberts, J.~Rodr\'\i{}guez-Quintero and J.~Segovia,
Phys. Rev. D \textbf{99}, no.3, 034013 (2019)
doi:10.1103/PhysRevD.99.034013
[arXiv:1811.08440 [nucl-th]].

\bibitem{Bednar:2018htv}
K.~D.~Bednar, I.~C.~Clo\"et and P.~C.~Tandy,
Phys. Lett. B \textbf{782}, 675-681 (2018)
doi:10.1016/j.physletb.2018.06.020
[arXiv:1803.03656 [nucl-th]].

\bibitem{Qin:2019hgk}
S.~x.~Qin, C.~D.~Roberts and S.~M.~Schmidt,
Few Body Syst. \textbf{60}, no.2, 26 (2019)
doi:10.1007/s00601-019-1488-x
[arXiv:1902.00026 [nucl-th]].

\bibitem{Chen:2019fzn}
C.~Chen, G.~I.~Krein, C.~D.~Roberts, S.~M.~Schmidt and J.~Segovia,
Phys. Rev. D \textbf{100}, no.5, 054009 (2019)
doi:10.1103/PhysRevD.100.054009
[arXiv:1901.04305 [nucl-th]].

\bibitem{Podolsky:1942zz} 
  B.~Podolsky,
  Phys.\ Rev.\  {\bf 62}, 68 (1942).
  doi:10.1103/PhysRev.62.68

\bibitem{Podolsky:1944zz} 
  B.~Podolsky and C.~Kikuchi,
  Phys.\ Rev.\  {\bf 65}, 228 (1944).
  doi:10.1103/PhysRev.65.228
  
\bibitem{Bufalo:2012tt} 
  R.~Bufalo, B.~M.~Pimentel and G.~E.~R.~Zambrano,
  Phys.\ Rev.\ D {\bf 86}, 125023 (2012)
  doi:10.1103/PhysRevD.86.125023
  [arXiv:1212.3542 [hep-th]].

\bibitem{Galvao:1986yq} 
  C.~A.~P.~Galv\~ao and B.~M.~Pimentel Escobar,
  Can.\ J.\ Phys.\  {\bf 66}, 460 (1988).
  doi:10.1139/p88-075

\bibitem{Cuzinatto:2005zr} 
  R.~R.~Cuzinatto, C.~A.~M.~de Melo and P.~J.~Pompeia,
  Annals Phys.\  {\bf 322}, 1211 (2007)
  doi:10.1016/j.aop.2006.07.006
  [hep-th/0502052].

\bibitem{Bufalo:2010sb} 
  R.~Bufalo, B.~M.~Pimentel and G.~E.~R.~Zambrano,
  Phys.\ Rev.\ D {\bf 83}, 045007 (2011)
  doi:10.1103/PhysRevD.83.045007
  [arXiv:1008.3181 [hep-th]].

\bibitem{Ji:2019phv}
C.~R.~Ji, A.~T.~Suzuki, J.~H.~O.~Sales and R.~Thibes,
Eur. Phys. J. C \textbf{79}, no.10, 871 (2019)
doi:10.1140/epjc/s10052-019-7384-1
[arXiv:1902.07632 [hep-th]].

\bibitem{Fischer:2008uz}
C.~S.~Fischer, A.~Maas and J.~M.~Pawlowski,
Annals Phys. \textbf{324}, 2408-2437 (2009)
doi:10.1016/j.aop.2009.07.009
[arXiv:0810.1987 [hep-ph]].

\bibitem{Alkofer:2008jy}
R.~Alkofer, M.~Q.~Huber and K.~Schwenzer,
Phys. Rev. D \textbf{81}, 105010 (2010)
doi:10.1103/PhysRevD.81.105010
[arXiv:0801.2762 [hep-th]].

\bibitem{Dudal:2008sp}
D.~Dudal, J.~A.~Gracey, S.~P.~Sorella, N.~Vandersickel and H.~Verschelde,
Phys. Rev. D \textbf{78}, 065047 (2008)
doi:10.1103/PhysRevD.78.065047
[arXiv:0806.4348 [hep-th]].

\bibitem{Aguilar:2004sw}
A.~C.~Aguilar and A.~A.~Natale,
JHEP \textbf{08}, 057 (2004)
doi:10.1088/1126-6708/2004/08/057
[arXiv:hep-ph/0408254 [hep-ph]].

\bibitem{Aguilar:2008xm}
A.~C.~Aguilar, D.~Binosi and J.~Papavassiliou,
Phys. Rev. D \textbf{78}, 025010 (2008)
doi:10.1103/PhysRevD.78.025010
[arXiv:0802.1870 [hep-ph]].

\bibitem{Aguilar:2012rz}
A.~C.~Aguilar, D.~Binosi and J.~Papavassiliou,
Phys. Rev. D \textbf{86}, 014032 (2012)
doi:10.1103/PhysRevD.86.014032
[arXiv:1204.3868 [hep-ph]].

\bibitem{Cucchieri:2007md}
A.~Cucchieri and T.~Mendes,
PoS \textbf{LATTICE2007}, 297 (2007)
doi:10.22323/1.042.0297
[arXiv:0710.0412 [hep-lat]].

\bibitem{Cucchieri:2007rg}
A.~Cucchieri and T.~Mendes,
Phys. Rev. Lett. \textbf{100}, 241601 (2008)
doi:10.1103/PhysRevLett.100.241601
[arXiv:0712.3517 [hep-lat]].

\bibitem{Oliveira:2008uf}
O.~Oliveira and P.~J.~Silva,
Phys. Rev. D \textbf{79}, 031501 (2009)
doi:10.1103/PhysRevD.79.031501
[arXiv:0809.0258 [hep-lat]].

\bibitem{Pennington:2011xs}
M.~R.~Pennington and D.~J.~Wilson,
Phys. Rev. D \textbf{84}, 119901 (2011)
doi:10.1103/PhysRevD.84.094028
[arXiv:1109.2117 [hep-ph]].

\bibitem{Oliveira:2012eh}
O.~Oliveira and P.~J.~Silva,
Phys. Rev. D \textbf{86}, 114513 (2012)
doi:10.1103/PhysRevD.86.114513
[arXiv:1207.3029 [hep-lat]].

\bibitem{Bogolubsky:2009dc}
I.~L.~Bogolubsky, E.~M.~Ilgenfritz, M.~M\"uller-Preussker and A.~Sternbeck,
Phys. Lett. B \textbf{676}, 69-73 (2009)
doi:10.1016/j.physletb.2009.04.076
[arXiv:0901.0736 [hep-lat]].

\bibitem{Ayala:2012pb}
A.~Ayala, A.~Bashir, D.~Binosi, M.~Cristoforetti and J.~Rodr\'iguez-Quintero,
Phys. Rev. D \textbf{86}, 074512 (2012)
doi:10.1103/PhysRevD.86.074512
[arXiv:1208.0795 [hep-ph]].

\bibitem{Strauss:2012dg}
S.~Strauss, C.~S.~Fischer and C.~Kellermann,
Phys. Rev. Lett. \textbf{109}, 252001 (2012)
doi:10.1103/PhysRevLett.109.252001
[arXiv:1208.6239 [hep-ph]].

\bibitem{Huber:2015ria}
M.~Q.~Huber,
Phys. Rev. D \textbf{91}, no.8, 085018 (2015)
doi:10.1103/PhysRevD.91.085018
[arXiv:1502.04057 [hep-ph]].

\bibitem{Cyrol:2016tym}
A.~K.~Cyrol, L.~Fister, M.~Mitter, J.~M.~Pawlowski and N.~Strodthoff,
Phys. Rev. D \textbf{94}, no.5, 054005 (2016)
doi:10.1103/PhysRevD.94.054005
[arXiv:1605.01856 [hep-ph]].

\bibitem{Boucaud:2018xup}
P.~Boucaud, F.~De Soto, K.~Raya, J.~Rodr\'iguez-Quintero and S.~Zafeiropoulos,
Phys. Rev. D \textbf{98}, no.11, 114515 (2018)
doi:10.1103/PhysRevD.98.114515
[arXiv:1809.05776 [hep-ph]].

\bibitem{Mintz:2018hhx}
B.~W.~Mintz, L.~F.~Palhares, G.~Peruzzo and S.~P.~Sorella,
Phys. Rev. D \textbf{99}, no.3, 034002 (2019)
doi:10.1103/PhysRevD.99.034002
[arXiv:1812.03166 [hep-th]].

\bibitem{Dudal:2018cli}
D.~Dudal, O.~Oliveira and P.~J.~Silva,
Annals Phys. \textbf{397}, 351-364 (2018)
doi:10.1016/j.aop.2018.08.019
[arXiv:1803.02281 [hep-lat]].

\bibitem{Aguilar:2019uob}
A.~C.~Aguilar, F.~De Soto, M.~N.~Ferreira, J.~Papavassiliou, J.~Rodr\'\i{}guez-Quintero and S.~Zafeiropoulos,
Eur. Phys. J. C \textbf{80}, no.2, 154 (2020)
doi:10.1140/epjc/s10052-020-7741-0
[arXiv:1912.12086 [hep-ph]].

\bibitem{Gunkel:2019xnh}
P.~J.~Gunkel, C.~S.~Fischer and P.~Isserstedt,
Eur. Phys. J. A \textbf{55}, no.9, 169 (2019)
doi:10.1140/epja/i2019-12868-1
[arXiv:1907.08110 [hep-ph]].

\bibitem{Gunkel:2020wcl}
P.~J.~Gunkel and C.~S.~Fischer,
[arXiv:2012.01957 [hep-ph]].

\bibitem{Huber:2020keu}
M.~Q.~Huber,
Phys. Rev. D \textbf{101}, 114009 (2020)
doi:10.1103/PhysRevD.101.114009
[arXiv:2003.13703 [hep-ph]].

\bibitem{Fischer:2005en}
C.~S.~Fischer, P.~Watson and W.~Cassing,
Phys. Rev. D \textbf{72} (2005), 094025
doi:10.1103/PhysRevD.72.094025
[arXiv:hep-ph/0509213 [hep-ph]].

\bibitem{Krassnigg:2009gd}
A.~Krassnigg,
PoS \textbf{CONFINEMENT8} (2008), 075
doi:10.22323/1.077.0075
[arXiv:0812.3073 [nucl-th]].

\bibitem{Aguilar:2009nf}
A.~C.~Aguilar, D.~Binosi, J.~Papavassiliou and J.~Rodr\'iguez-Quintero,
Phys. Rev. D \textbf{80}, 085018 (2009)
doi:10.1103/PhysRevD.80.085018
[arXiv:0906.2633 [hep-ph]].

\bibitem{Curci:1976bt}
G.~Curci and R.~Ferrari,
Nuovo Cim. A \textbf{32} (1976), 151-168
doi:10.1007/BF02729999

\bibitem{Alkofer:2008tt}
R.~Alkofer, C.~S.~Fischer, F.~J.~Llanes-Estrada and K.~Schwenzer,
Annals Phys. \textbf{324}, 106-172 (2009)
doi:10.1016/j.aop.2008.07.001
[arXiv:0804.3042 [hep-ph]].

\bibitem{Kizilersu:2009kg}
A.~Kizilersu and M.~R.~Pennington,
Phys. Rev. D \textbf{79}, 125020 (2009)
doi:10.1103/PhysRevD.79.125020
[arXiv:0904.3483 [hep-th]].

\bibitem{Bashir:2011dp}
A.~Bashir, R.~Berm\'udez, L.~Chang and C.~D.~Roberts,
Phys. Rev. C \textbf{85}, 045205 (2012)
doi:10.1103/PhysRevC.85.045205
[arXiv:1112.4847 [nucl-th]].

\bibitem{Rojas:2013tza}
E.~Rojas, J.~P.~B.~C.~de Melo, B.~El-Bennich, O.~Oliveira and T.~Frederico,
JHEP \textbf{10}, 193 (2013)
doi:10.1007/JHEP10(2013)193
[arXiv:1306.3022 [hep-ph]].

\bibitem{Rojas:2014tya} 
  E.~Rojas, B.~El-Bennich, J.~P.~B.~C.~De Melo and M.~A.~Paracha,
  Few Body Syst.\  {\bf 56}, no. 6-9, 639 (2015)
  doi:10.1007/s00601-015-1020-x
  [arXiv:1409.8620 [hep-ph]].
  
\bibitem{Aguilar:2014lha}
A.~C.~Aguilar, D.~Binosi, D.~Iba\~nez and J.~Papavassiliou,
Phys. Rev. D \textbf{90}, no.6, 065027 (2014)
doi:10.1103/PhysRevD.90.065027
[arXiv:1405.3506 [hep-ph]].

\bibitem{Williams:2014iea}
R.~Williams,
Eur. Phys. J. A \textbf{51}, no.5, 57 (2015)
doi:10.1140/epja/i2015-15057-4
[arXiv:1404.2545 [hep-ph]].

\bibitem{Pelaez:2015tba}
M.~Pel\'aez, M.~Tissier and N.~Wschebor,
Phys. Rev. D \textbf{92}, no.4, 045012 (2015)
doi:10.1103/PhysRevD.92.045012
[arXiv:1504.05157 [hep-th]].

\bibitem{Pennington:2016vxv}
S.~Jia and M.~R.~Pennington,
Phys. Rev. D \textbf{94}, no.11, 116004 (2016)
doi:10.1103/PhysRevD.94.116004
[arXiv:1610.10049 [nucl-th]].

\bibitem{Williams:2015cvx}
R.~Williams, C.~S.~Fischer and W.~Heupel,
Phys. Rev. D \textbf{93}, no.3, 034026 (2016)
doi:10.1103/PhysRevD.93.034026
[arXiv:1512.00455 [hep-ph]].

\bibitem{Sternbeck:2017ntv}
A.~Sternbeck, P.~H.~Balduf, A.~K\i{}z\i{}lersu, O.~Oliveira, P.~J.~Silva, J.~I.~Skullerud and A.~G.~Williams,
PoS \textbf{LATTICE2016}, 349 (2017)
doi:10.22323/1.256.0349
[arXiv:1702.00612 [hep-lat]].

\bibitem{Aguilar:2018epe}
A.~C.~Aguilar, J.~C.~Cardona, M.~N.~Ferreira and J.~Papavassiliou,
Phys. Rev. D \textbf{98}, no.1, 014002 (2018)
doi:10.1103/PhysRevD.98.014002
[arXiv:1804.04229 [hep-ph]].

\bibitem{Serna:2018dwk}
F.~E.~Serna, C.~Chen and B.~El-Bennich,
Phys. Rev. D \textbf{99}, no.9, 094027 (2019)
doi:10.1103/PhysRevD.99.094027
[arXiv:1812.01096 [hep-ph]].

\bibitem{Albino:2018ncl}
L.~Albino, A.~Bashir, L.~X.~G.~Guerrero, B.~E.~Bennich and E.~Rojas,
Phys. Rev. D \textbf{100}, no.5, 054028 (2019)
doi:10.1103/PhysRevD.100.054028
[arXiv:1812.02280 [nucl-th]].

\bibitem{Oliveira:2018ukh}
O.~Oliveira, W.~de Paula, T.~Frederico and J.~P.~B.~C.~de Melo,
Eur. Phys. J. C \textbf{79}, no.2, 116 (2019)
doi:10.1140/epjc/s10052-019-6617-7
[arXiv:1807.10348 [hep-ph]].

\bibitem{Oliveira:2020yac}
O.~Oliveira, T.~Frederico and W.~de Paula,
Eur. Phys. J. C \textbf{80}, no.5, 484 (2020)
doi:10.1140/epjc/s10052-020-8037-0
[arXiv:2006.04982 [hep-ph]].

\bibitem{Gao:2020qsj}
F.~Gao and J.~M.~Pawlowski,
Phys. Rev. D \textbf{102}, no.3, 034027 (2020)
doi:10.1103/PhysRevD.102.034027
[arXiv:2002.07500 [hep-ph]].

\bibitem{Nakanishi:1965zza} 
  N.~Nakanishi,
  Phys.\ Rev.\  {\bf 138}, B1182 (1965).
  doi:10.1103/PhysRev.138.B1182

\bibitem{Fischer:2009jm} 
  C.~S.~Fischer and R.~Williams,
  Phys.\ Rev.\ Lett.\  {\bf 103}, 122001 (2009)
  doi:10.1103/PhysRevLett.103.122001
  [arXiv:0905.2291 [hep-ph]].

\bibitem{Tanabashi:2018oca} 
  M.~Tanabashi {\it et al.} [Particle Data Group],
  Phys.\ Rev.\ D {\bf 98}, no. 3, 030001 (2018).
  doi:10.1103/PhysRevD.98.030001

\bibitem{Davies:2010ip} 
  C.~T.~H.~Davies, C.~McNeile, E.~Follana, G.~P.~Lepage, H.~Na and J.~Shigemitsu,
  Phys.\ Rev.\ D {\bf 82}, 114504 (2010)
  doi:10.1103/PhysRevD.82.114504
  [arXiv:1008.4018 [hep-lat]].
 
\bibitem{Aoki:2019cca}
S.~Aoki \textit{et al.} [Flavour Lattice Averaging Group],
Eur. Phys. J. C \textbf{80} (2020) no.2, 113
doi:10.1140/epjc/s10052-019-7354-7
[arXiv:1902.08191 [hep-lat]].
  
\bibitem{deMelo:2014gea}
J.~P.~B.~C.~de Melo, K.~Tsushima, B.~El-Bennich, E.~Rojas and T.~Frederico,
Phys. Rev. C \textbf{90}, no.3, 035201 (2014)
doi:10.1103/PhysRevC.90.035201
[arXiv:1404.5873 [hep-ph]].



\end{thebibliography}
\end{document}